\documentclass[graybox]{svmult}[12pt]

\usepackage[margin=1in,footskip=0.25in]{geometry}

\usepackage{graphicx}
\usepackage{epstopdf}
\usepackage{subfigure}
\usepackage{tikz}
\usepackage{multicol}
\usepackage{float}
\usepackage{afterpage}

\usepackage{mathptmx}       
\usepackage{helvet}         
\usepackage{courier}        
\usepackage{makeidx}         
\usepackage{graphicx}        
\usepackage{multicol}        

\begin{document}

\title*{Fitting logistic multilevel models with crossed random effects via Bayesian Integrated Nested Laplace Approximations: a simulation study}
\titlerunning{Fitting logistic multilevel models with crossed random effects via INLA}
\author{L. Grilli \and F. Innocenti}
\institute{Leonardo Grilli:  Dipartimento di Statistica, Informatica, Applicazioni "G. Parenti", Universit\'a di Firenze \texttt{grilli@disia.unifi.it} \and Francesco Innocenti: \texttt{f.innocenti91@gmail.com}}

\maketitle

\abstract{Fitting cross-classified multilevel models with binary response is challenging. In this setting a promising method is Bayesian inference through Integrated Nested Laplace Approximations (INLA), which performs well in several latent variable models. Therefore we devise a systematic simulation study to assess the performance of INLA with cross-classified logistic data under different scenarios defined by the magnitude of the random effects variances, the number of observations, the number of clusters, and the degree of cross-classification. In the simulations INLA is systematically compared with the popular method of Maximum Likelihood via Laplace Approximation. By an application to the classical salamander mating data, we compare INLA with the best performing methods. Given the computational speed and the generally good performance, INLA turns out to be a valuable method for fitting the considered cross-classified models.}

\keywords{
binary response; crossed random effects; Generalized Linear Mixed Models; INLA; Maximum Likelihood via Laplace Approximation; non-hierarchical data; random effects; salamander mating data.}

\section{Introduction}
Cross-classified data are non-hierarchical structures where lower level units belong to pairs or combinations of higher level units formed by crossing each other two or more higher level factors \cite{gold94,leckie13}. Examples include children cross-classified by primary and secondary schools \cite{mmmc} or by school and neighborhood \cite{rg}, and responses nested in the combination of test items and persons \cite{vannoor}. There are different degrees of cross-classification, that can be categorized essentially into two types \cite{luokwok}:
\begin{itemize}
\item complete cross-classification: units in a cluster of one factor belong to all the clusters of the other crossed factor, and vice versa;
\item partial cross-classification: units in a cluster of one factor belong to a subset of the clusters of the other crossed factor.
\end{itemize}

Linear cross-classified models have been widely studied in literature \cite{luokwok,Meyers,shi} and the related estimation issues have been satisfactorily addressed \cite{clay, rg}, as testified by the large number of published applications (see \cite{fielgol} for a detailed review). 

On the other hand, fitting logistic cross-classified models is difficult for two reasons: (\textit{i}) the distribution of the response conditional on the random effects is Bernoulli, thus the marginal likelihood is not in closed form; (\textit{ii}) due to the cross-classification of the random effects, the variance-covariance matrix is not block-diagonal. In the simpler case of nested random effects (Generalized Linear Mixed Models), several methods are available to obtain Maximum Likelihood (ML) estimates, including linearization (MQL \cite{gold91}, PQL \cite{breslow}) and numerical integration, such as Laplace Approximation (MLLA) \cite{rau} and Adaptive Gaussian Quadrature (AGQ) \cite{rabe2005}). In general, maximum likelihood methods tend to underestimate the variance components, especially in settings with a small number of clusters \cite{joe}. In the challenging case of crossed random effects, the above methods can still be used, even if better performances can be obtained by special algorithms based on data augmentation, such Monte Carlo Expectation Maximization (MCEM) \cite{McCull} and Alternating Imputation Posterior (AIP) estimation \cite{clay,cho}.

Bayesian methods generally have a better performance in complex random effects models \cite{browdrap}. However, the standard Bayesian method, namely MCMC \cite{zeger}, has some practical limitations because of the computational burden and the difficulties in assessing convergence. 
A possible solution is represented by INLA, namely Integrated Nested Laplace Approximations \cite{inla}: indeed, INLA directly approximates the posterior distribution, thus avoiding complex simulation-based methods. INLA is promising because of the good performance observed in logistic models with nested random effects \cite{grilli}, where it is fast (nearly as frequentist quadrature methods) and accurate (slightly more than MCMC). Therefore, it is worth to investigate the performance of INLA in logistic cross-classified multilevel models. In this setting, the only application we are aware of is reported in the Supplementary Material of \cite{fong}, where INLA is used to fit model C of \cite{zeger} on the classical salamander mating data. In that instance, INLA seems to underestimate the variance components, but a comprehensive evaluation of the method requires a systematic simulation study. We therefore devise a simulation study to evaluate INLA for a logistic model with two crossed random effects under several scenarios. The results are compared with those obtained with Maximum Likelihood via Laplace Approximation (MLLA), which is the default choice in many programs and it is similar to INLA in terms of computational time. We do not consider AGQ, which is generally superior to Laplace Approximation, but it turns out to be infeasible in some scenarios.
In the simulation study, we devote particular attention to situations with a small number of clusters, different degrees of cross-classification and tiny random effects variances.
Furthermore, in order to compare INLA with a wide set of estimation methods (MCMC, MCEM, AIP), we apply it to the classical salamander mating data \cite{salamander,cho,zeger}.

The rest of the paper is organized as follows. In Section \ref{design} the simulation design is described. In Section \ref{results} the findings of the simulation study are commented, considering INLA with two alternative prior distributions and MLLA. In Section \ref{salam} INLA is applied to the salamander mating data, allowing a comparison with several efficient algorithms. Section \ref{conclusion} offers some final remarks.
The Supplementary Material collects further simulation results not reported in the paper.

\section{The simulation design} 
\label{design}
\subsection{Model and sample structure}
\label{sub:scenarios}
We consider a random intercept logistic model with two crossed random effects. Let $Y_{i(j_{1}j_{2})}$ be a Bernoulli random variable for level 1 unit $i$ (e.g. student) nested in two crossed classifications at level 2 (e.g. school and neighborhood) with $ j_{1}= 1,..., N_{1}$ and $ j_{2}= 1,..., N_{2}$. 
Defining $\pi_{i(j_{1}j_{2})}=P(Y_{i(j_{1}j_{2})}=1\vert x_{1i(j_{1}j_{2})}, x_{2i(j_{1}j_{2})}, z_{j_{1}}, z_{j_{2}}, u_{j_{1}}, u_{j_{2}})$, the considered model is:
\begin{equation} 
\label{modello}
logit(\pi_{i(j_{1}j_{2})})=\alpha+\beta_{1}x_{1i(j_{1}j_{2})}+\beta_{2}x_{2i(j_{1}j_{2})}+\gamma_{1}z_{1j_{1}}+\gamma_{2}z_{2j_{2}}+u_{j_{1}}+u_{j_{2}}
\end{equation}
\[u_{j_{1}}\sim N(0,\sigma^{2}_{u_{j_{1}}})\;\;\; u_{j_{2}}\sim N(0,\sigma^{2}_{u_{j_{2}}})\]\\
where $x_{1i(j_{1}j_{2})}$ is a continuous level 1 variable, $x_{2i(j_{1}j_{2})}$ is a binary level 1 variable, $z_{1j_{1}}$ and $z_{2j_{2}}$ are binary level 2 variables (the former related to classification 1, the latter related to classification 2). The continuous covariate $x_{1i(j_{1}j_{2})}$ is drawn from a standard Normal distribution, whereas the binary covariates are drawn from independent Bernoulli distributions with success probability equal to $0.5$.

The true values of parameters in model (\ref{modello}) are set as follows: $\alpha=0.1$ (so that the baseline individual has a probability of success slightly greater than $0.5$), $\beta_{1}=0.1$, and $\beta_{2}=\gamma_{1}=\gamma_{2}=0.4$. Given that in a standard Normal distribution about $95\%$ of the probability lies between $-2$ and $+2$, setting $\beta_{1}=0.1$ ensures that the continuous covariate has an effect comparable to that of the binary covariates. The values of the regression coefficients are constant across all configurations.

On the other hand, several values are considered for the variances of the random effects $\sigma^{2}_{u_{j_{1}}}$ and $\sigma^{2}_{u_{j_{2}}}$ since it is known that they strongly affect the performance of the estimation methods and the importance of the prior distribution \cite{browdrap,gel}.
Specifically, we consider four configurations yielded by setting the variances of the two random effects at either $0.01$ (low impact of the random effects) or $0.25$ (sizeable impact of the random effects). To see the impact of the random effects, note that a variance $\sigma^{2}_{u_{j_{.}}}=0.01$ corresponds to a standard deviation $\sigma_{u_{j_{.}}}=0.1$, thus under normality the random effect approximately has $95\%$ probability of lying in the interval $[-0.2,0.2]$, corresponding to the central interval $[0.45,0.55]$ in terms of probability of the response variable. Similarly, $\sigma^{2}_{u_{j_{.}}}=0.25$ implies $\sigma_{u_{j_{.}}}=0.50$ so that the central $95\%$ interval of the probability is $[0.27,0.73]$. In the Supplementary Material we also report simulations with both variances at $1.00$, though we do not put much emphasis on the results since random effects of such size are rarely found in applications. Anyway, it is worth to note that inference about random effects with variance $1.00$ is not problematic and, indeed, the performances of all the considered estimators are satisfactory.

In order to evaluate the influence of the degree of cross-classification on the performance of INLA and MLLA \cite{luokwok}, we consider three scenarios ranging from complete cross-classification to an almost hierarchical structure:
\begin{itemize}
\item a complete cross-classified structure: we consider a square cross-classification matrix, namely the two classification factors have the same number of clusters $N_{1}=N_{2}$. Since Bayesian and frequentist methods can differ substantially in scenarios with a small number of clusters, we focus our investigation on the case $N_{1}=N_{2}=10$, considering four different values for the number of observations per cell $n$ (1, 5, 10, 20). Moreover, in order to assess the asymptotic behaviour of the estimators, we consider four different values for $N_{1}=N_{2}$ (10, 20, 50, 80), with $n=10$ observations per cell in each scenario. 
\item two partial cross-classified structures: we generate those structures as follows \cite{luokwok}:
\begin{enumerate}
\item Generate a hierarchical three-level model: classification 1 is the third level with 10 clusters, classification 2 is the second level with 10 clusters within each third-level unit, and 100 observations are nested within each second-level unit.
\item Randomly draw 10 second-level units (called \textit{feeders}).
\item Randomly draw $k$ third-level units (called \textit{receivers}), where $k$ is set to either 2 or 5.
\item For each feeder, randomly assign the observations to the receivers (50 observations per receiver in case of 2 receivers, and 20 observations per receiver in case of 5 receivers)
\end{enumerate}
\end{itemize}
Table \ref{table:receiver1} represents a partial cross-classified structure with 10 feeders and 5 receivers.
Note that the three structures outlined above have the same sample size (1000 observations), but they differ in the distribution of the empty cells (which is an indicator of the degree of cross-classification \cite{luokwok}) and in the number of observations per cell ($n=10$ in the complete cross-classified structure, $n=50$ in the structure with 2 receivers, and $n=20$ in structure with 5 receivers).
\begin{table}
\caption{Partial cross-classified structure with 10 feeders and 5 receivers (the symbol x denotes the presence of at least one observation).}\label{table:receiver1}
\centering
\begin{tabular}{cccccccccccc}
\multicolumn{12}{c}{RECEIVERS}   \\ \hline
 & & 1& 2& 3& 4& 5& 6& 7& 8& 9&10\\ \hline
 & 1&  & x& x& &x & &x &  & &x  \\
F& 2&  & x& x& &  & &x &  & x&x \\
E& 3& x&x &x &x&  & &  &x & &   \\
E& 4&  &x & x& &x &x&  &  & &x  \\
D& 5& x&  &  & &x &x&  &x &x&   \\
E& 6&  &x &x &x&x & &  &  &x&   \\
R& 7&  &x &x & &x & &x & x& &    \\
S& 8&  & x&  &x& x& x& x  & & & \\
 & 9& x&  & x&x&  & x& &  & &x  \\
 &10&  & x&  & & x& x& x& & &x  \\
\hline
\end{tabular}
\end{table}

\subsection{Estimation methods and prior distributions}
\label{sub:priors}
The simulation study is performed with the following packages implemented in the \texttt{R} software (version 3.2.2):
\begin{itemize}
\item the \texttt{inla} package (version 0.0-1440400394) for INLA - Bayesian inference through Integrated Nested Laplace Approximations \cite{rue,reviewINLA};
\item the \texttt{lme4} package (version 1.1-9) for MLLA - Maximum Likelihood via Laplace Approximation \cite{lme4}.
\end{itemize}
In the INLA algorithm, we rely on the default method (Simplified Laplace Approximation) since preliminary simulations showed that its accuracy is similar to the alternative, more complex method.

\begin{figure}
\centering{ \includegraphics[scale=0.3]{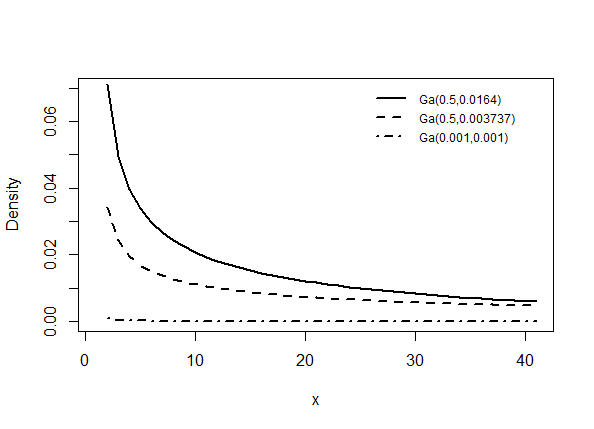}}
\caption{Probability densities of the considered prior distributions in $[0,40]$.}
\label{figure:priors}
\end{figure}

In Bayesian inference with INLA we specify a Normal prior distribution with zero mean and large variance for the regression coefficients (the default of the \texttt{inla} function). Since we focus on scenarios with small numbers of clusters and variances close to zero, the choice of the prior distribution for the variance components is crucial. As usual in Bayesian software, the \texttt{inla} function allows us to specify the prior distribution of the precision, instead of the variance. 
We avoid the default gamma prior Ga(1, 0.0005) because it has a poor performance in logistic models with nested random effects \cite{grilli}.
For the simulation study we choose two alternative priors for the precisions: Ga(0.001,0.001), namely the standard choice in the popular BUGS software, and Ga(0.5, 0.003737) specified according to the criterion proposed by Fong et al. \cite{fong}, which consists in setting the parameters of the Gamma in order to obtain a given marginal distribution for the random effects. In particular, Ga(0.5, 0.003737) yields a marginal Cauchy distribution having 95\% probability of $e^{u_{j_{.}}}\in \left[\frac{1}{3},3\right]$, corresponding to a central interval for the probability of the response variable equal to $[0.25,0.75]$. It is worth to note that the selected prior Ga(0.5, 0.003737) is different from the prior usually derived by applying the Fong et. al criterion (like in the simulations of \cite{grilli}), which is Ga(0.5, 0.0164). This prior amounts to random effects with a stronger impact, as it yields a marginal Cauchy distribution having 95\% probability of $e^{u_{j_{.}}}\in \left[\frac{1}{10},10\right]$, corresponding to a central interval for the probability of the response variable equal to $[0.09,0.91]$. In the simulations we tried both the priors derived by the Fong et. al criterion, but we retained only Ga(0.5, 0.003737) as it was outperforming the other one. The densities of the three mentioned priors are depicted in Figure \ref{figure:priors}, showing that Ga(0.5, 0.003737) is less informative than Ga(0.5, 0.0164), though more informative than Ga(0.001,0.001).

\section{Simulation results}
\label{results}
\subsection{Measures of performance} \label{sub:measure}
The performances of INLA and MLLA are compared on the basis of the following measures of accuracy \cite{luokwok,Meyers,shi}, where $m$ is one of the scenarios defined in Subsection \ref{sub:scenarios} and $l$ is one of the $L$ Monte Carlo replicates:
\begin{itemize}
\item relative bias for the estimates of the regression coefficients and variance components: \[RB(\hat{\theta}_{m})=\frac{\hat{\theta}_{m}-\theta}{\theta},\]
where $\hat{\theta}_{m}=\frac{\sum_{l=1}^{L}\hat{\theta}_{lm}}{L}$ is the Monte Carlo average of the estimates $\hat{\theta}_{lm}$ (point estimates for MLLA and posterior means for INLA) and $\theta$ is the population parameter;
\item relative bias for the standard errors of the regression coefficients: \[RB(S_{\hat{\theta}_{m}})=\frac{\overline{SE}(\hat{\theta}_{m})-SD(\hat{\theta}_{m})}{SD(\hat{\theta}_{m})},\]
where $\overline{SE}(\hat{\theta}_{m})=\frac{\sum_{l=1}^{L} SE(\hat{\theta}_{lm})}{L}$ is the Monte Carlo average standard error and $SD(\hat{\theta}_{m})$ is the Monte Carlo standard error, namely the standard deviation of the $L$ estimates $\hat{\theta}_{lm}$.
\end{itemize}
The standard errors of the variance components are not considered because they are not provided by the \texttt{glmer} function in the \texttt{lme4} package (in general, it is not advisable to exploit the standard errors to make inference on the variance components).

\subsection{Extreme estimates of the variance components} \label{sub:extreme}
In order to give practical advice to applied researchers, it is worth to study when the two considered estimators yield extreme values for the estimates of the variance components. For MLLA it sometimes happens that $\hat{\sigma}_{u_{j.}}^2=0$, namely the estimate is on the border of the parameter space. This problem does not occur with INLA since the priors push the estimates into the parameter space; however, INLA sometimes yield unrealistically large estimates. In the following, we label as aberrant the estimates larger than 2, namely $\hat{\sigma}_{u_{j.}}^2 > 2$. Such threshold is necessarily subjective as it corresponds to the largest value that a researcher is willing to trust. For each scenario we report the percentages of null estimates of MLLA and aberrant estimates of INLA out of the 500 replicates (see Tables \ref{table:anomalie1}, \ref{table:anomalie2} and \ref{table:anomalie3}).

In Bayesian inference with non-informative priors, the usual action in case of aberrant estimates is to change the priors, therefore we discard the replicates where at least one of the variance components is larger than 2 and compute the relative bias on the remaining replicates $L \le 500$ (note that in most scenarios there are no aberrant estimates, thus $L = 500$). Discarding the replicates with aberrant estimates has a noticeable effect on the Monte Carlo relative bias of the variance components, whereas the effect on the regression coefficients is negligible.

\subsection{Scenarios with few clusters}
In our simulation study we devote particular attention to scenarios with few clusters because in these cases the estimation of the variances of the random effects is challenging and the influence of prior distributions is amplified, so that Bayesian and Maximum Likelihood methods may yield considerably different results \cite{browdrap}. 

In Table \ref{table:complete0505} we compare the relative biases (net of aberrant estimates) for the regression coefficients yielded by INLA and MLLA in a complete $10\times 10$ cross-classification matrix with varying number of observations per cell. Note that INLA and MLLA give similar results for the regression coefficients: both methods yield relative biases smaller than 10\% even with $n=5$ observations per cell and they decrease for larger cell sample sizes. However, the direction of the biases is hardly predictable. On the other hand, INLA and MLLA differ in the estimation of the standard errors of the regression coefficients: INLA yields more accurate standard errors for larger values of the random effects variances $\sigma_{j_{1}}^{2}$ and $\sigma_{j_{2}}^{2}$, whereas in this regard MLLA performs better when $\sigma_{j_{1}}^{2}$ and $\sigma_{j_{2}}^{2}$ are close to zero.

\begin{table}
\caption{Relative bias for regression coefficients (relative bias of standard errors in parenthesis). Logistic model of equation (\ref{modello}) with $\sigma_{j_{1}}^{2}=\sigma_{j_{2}}^{2}=0.25$. Complete cross-classification with $N_{1}=N_{2}=10$ and varying number of observations per cell $n$.}\label{table:complete0505}
\centering
\begin{tabular}{|c|ccc|}
\hline
$n$ & INLA Ga(0.001,0.001)& INLA Ga(0.5,0.003737)& MLLA\\
\hline
    & & $\alpha$& \\ \cline{3-3}
1   &  -0.118 (0.014)&  0.167 (-0.148)& -0.060 (-0.092)\\
5   &  0.078 (0.174)&  0.380 (-0.020)&  0.056 (-0.023)\\
10  & -0.110 (0.102)& -0.030 (0.003)& -0.120 (-0.103)\\
20  & -0.110 (0.002)& 0.120 (-0.007)& -0.120 (-0.153)\\
    & & $\beta_{1}$& \\\cline{3-3}
1   &0.364 (-0.141)&  0.249 (-0.120)&  0.270 (-0.078)\\
5   &0.000 (0.000)& 0.060 (-0.081)& -0.014 (0.016)\\
10  &0.050 (-0.018)&  0.060 (-0.047)&  0.040 (-0.012)\\
20  & 0.030 (0.016)&  -0.010 (-0.037)&  0.020 (0.018)\\
    & & $\beta_{2}$& \\\cline{3-3}
1   & 0.113 (-0.081)&  -0.057 (0.010)&  0.035 (-0.011)\\
5   &0.064 (0.015)&  0.017 (-0.021)&  0.050 (0.031)\\
10  & 0.025 (-0.006)& 0.003 (0.065)&  0.017 (0.008)\\
20  & -0.003 (-0.038)& -0.003 (-0.024)& -0.005 (-0.036)\\
    & & $\gamma_{1}$& \\\cline{3-3}
1   & 0.220 (-0.003)&  0.110 (-0.149)&  0.092 (-0.081)\\
5   & 0.022 (0.120)&   -0.105 (-0.044)&  0.002 (-0.082)\\
10  &0.008 (0.041)&  0.080  (-0.009)&  0.000 (-0.191)\\
20  &-0.018 (0.069)& 0.013 (0.002)& -0.020 (-0.111)\\
    & & $\gamma_{2}$& \\\cline{3-3}
1   & 0.230 (-0.062)&  0.106 (-0.218)&  0.142 (-0.148)\\
5   & -0.053 (-0.132)&  -0.030 (-0.089)&  -0.064 (-0.158)\\
10  &-0.008 (0.040)&   0.065 (-0.025)&  -0.013 (-0.193)\\
20  &-0.005 (-0.013)& -0.018 (-0.015)& -0.008 (-0.022)\\
\hline
\end{tabular}
\end{table}

\begin{figure}
\centerline{\includegraphics[scale=0.3]{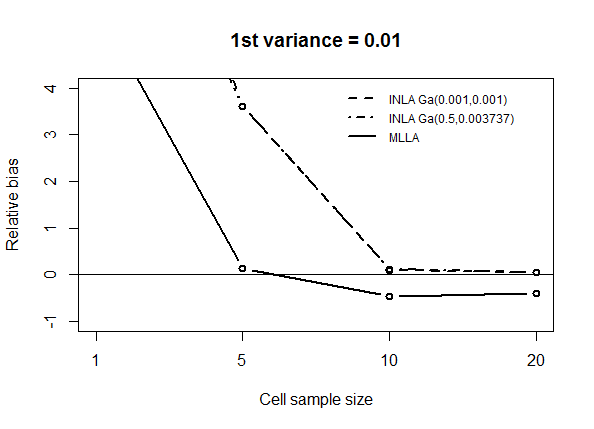} 
            \includegraphics[scale=0.3]{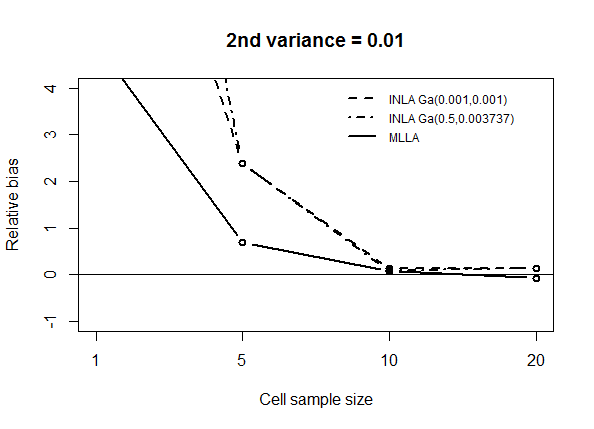}}
\centerline{\includegraphics[scale=0.3]{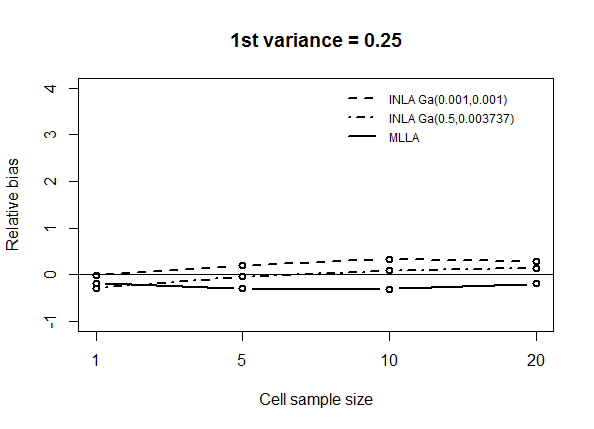}
            \includegraphics[scale=0.3]{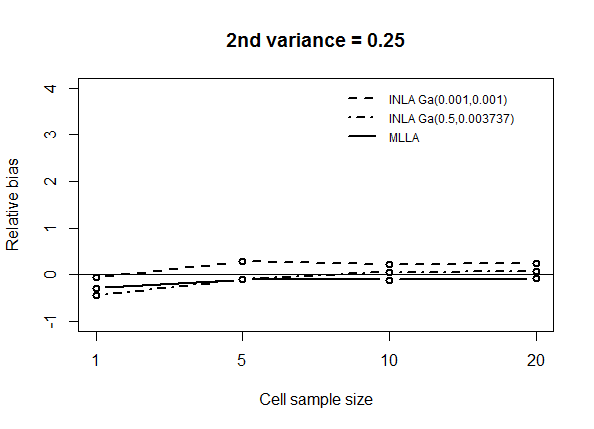}}
\centerline{\includegraphics[scale=0.3]{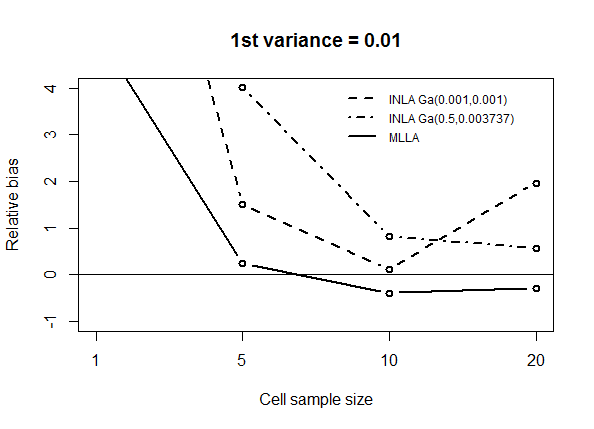}
            \includegraphics[scale=0.3]{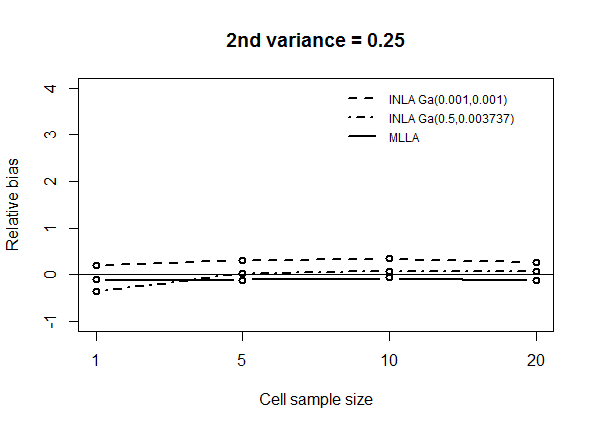} }
\centerline{\includegraphics[scale=0.3]{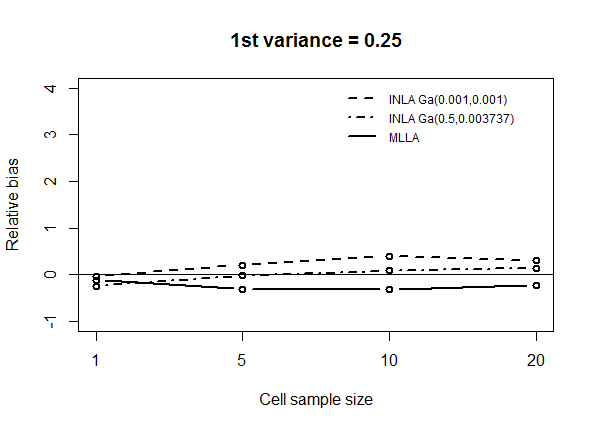}
            \includegraphics[scale=0.3]{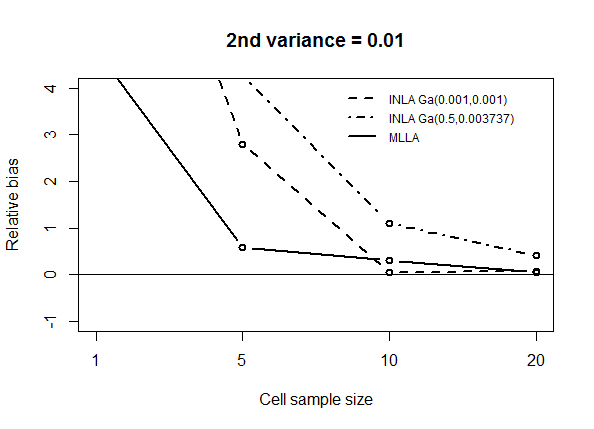} }
\caption{Relative bias for the variance components of the logistic model of equation (\ref{modello}). Complete cross-classification with $N_{1}=N_{2}=10$ and varying number of observations per cell $n$. Each pair of graphs corresponds to a combination of random effects variances ($\sigma_{j_{1}}^{2}, \sigma_{j_{2}}^{2}$): (0.01, 0.01), (0.25, 0.25), (0.01, 0.25), (0.25, 0.01).}
\label{figure:fewcluster}
\end{figure}

Figure \ref{figure:fewcluster} reports the relative biases (net of aberrant estimates) for random effects variances, highlighting the differences among MLLA and INLA with the two priors. When both variance components are 0.01, all the methods overestimate the population values, but the biases rapidly decline as the cell sample size $n$ increases ($n=10$ is enough for INLA, though not for MLLA as it underestimates the first variance). The two priors yield similar results. When both variance components are 0.25, the three methods perform well even for small cell sample sizes. Note that, generally, MLLA underestimates the variance components, whereas INLA overestimates it, with the prior Ga(0.5,0.003737) outperforming Ga(0.001,0.001). The cases where the variance components have markedly different sizes, namely 0.01 and 0.25 and viceversa, are troubling since the low variance component can be badly estimated even for $n=10$ or $n=20$. It is worth to note that the last two configurations have been considered as they are especially challenging, though they are unlikely in practice. In the configuration with variance components $\sigma_{j_{1}}^{2}=0.01$ and $\sigma_{j_{2}}^{2}=0.25$, INLA with the prior Ga(0.001,0.001) shows an anomalous behaviour because the bias abruptly increases when moving from cell size $n=10$ to $n=20$. This pattern is analyzed in detail in the Supplementary Material.

As discussed at the end of Subsection \ref{sub:extreme}, the considered estimation methods are prone to different kinds of extreme estimates for the variance components, namely MLLA may yield zero values, whereas INLA may yield very large values (here considered to be aberrant when larger than 2). Table \ref{table:anomalie1} reports the percentage of extreme estimates out of 500 for each scenario. For MLLA the issue of zero estimates is severe (above 20\%) for low variance components even in large sample sizes, whereas it is severe for variance components at 0.25 only in scenarios with cell size $n=1$. The opposite issue for INLA, namely aberrant estimates, has noticeable percentages only for variance components at 0.25 and cell size $n=1$. As expected, aberrant estimates are more likely with the more informative prior Ga(0.5,0.003737).

\begin{table}
\caption{Percentage of extreme estimates out of the 500 replicates. Complete cross-classification with $N_{1}=N_{2}=10$ and varying number of observations per cell $n$.} \label{table:anomalie1}
\centering
\begin{tabular}{|c|ccc|}
\hline
$n$ & INLA Ga(0.001,0.001)& INLA Ga(0.5,0.003737)& MLLA\\
\hline
    &\% $\hat{\sigma}_{j_{1}}^{2}> 2$ \& \% $\hat{\sigma}_{j_{2}}^{2}> 2$ &\% $\hat{\sigma}_{j_{1}}^{2}> 2$ \& \% $\hat{\sigma}_{j_{2}}^{2}> 2$ &\% $\hat{\sigma}_{j_{1}}^{2}=0$ \& \% $\hat{\sigma}_{j_{2}}^{2}=0$ \\\cline{2-4}
    & & $\sigma_{j_{1}}^{2}=\sigma_{j_{2}}^{2}=0.01$&\\\cline{3-3}
1   & 0.4 $\;\;\;\;\;\;$ 1.0&  5.4 $\;\;\;\;\;\;$    3.8  &      47.0   $\;\;\;\;\;\;$    54.2 \\
5   &  0.0 $\;\;\;\;\;\;$ 0.0   &   0.0   $\;\;\;\;\;\;$    0.0  &    46.6    $\;\;\;\;\;\;$     42.0\\
10  &   0.0   $\;\;\;\;\;\;$   0.0  &    0.0  $\;\;\;\;\;\;$     0.0 &     44.6   $\;\;\;\;\;\;$    35.6\\ 
20  &   0.0  $\;\;\;\;\;\;$    0.0  &    0.0   $\;\;\;\;\;\;$    0.0  &      34.0   $\;\;\;\;\;\;$      24.0\\ 
    & & $\sigma_{j_{1}}^{2}=\sigma_{j_{2}}^{2}=0.25$&\\\cline{3-3}
1   & 5.2   $\;\;\;\;\;\;$   4.0     &    11.0  $\;\;\;\;\;\;$   10.8   &   23.4  $\;\;\;\;\;\;$     24.8\\ 
5   & 0.0   $\;\;\;\;\;\;$   0.0 &     0.0   $\;\;\;\;\;\;$    0.0   &    3.2    $\;\;\;\;\;\;$    1.8\\ 
10  & 0.0  $\;\;\;\;\;\;$    0.0&      0.0   $\;\;\;\;\;\;$    0.0  &     0.8   $\;\;\;\;\;\;$       0.0\\ 
20  & 0.0  $\;\;\;\;\;\;$    0.0&      0.0    $\;\;\;\;\;\;$   0.0 &        0.0   $\;\;\;\;\;\;$       0.0\\ 
    & & $\sigma_{j_{1}}^{2}=0.01\;\;\;\sigma_{j_{2}}^{2}=0.25$&\\\cline{3-3}
1   &  0.6 $\;\;\;\;\;\;$   5.6  &  4.0  $\;\;\;\;\;\;$     12.0    &    38.0  $\;\;\;\;\;\;$     28.4\\
5   &0.0   $\;\;\;\;\;\;$   0.0   &   0.0   $\;\;\;\;\;\;$    0.0    &    29.0  $\;\;\;\;\;\;$      2.2\\
10  &0.0  $\;\;\;\;\;\;$    0.0 &     0.0   $\;\;\;\;\;\;$    0.0   &   39.4 $\;\;\;\;\;\;$         0.0\\ 
20  & 0.0  $\;\;\;\;\;\;$    0.0  &    0.0 $\;\;\;\;\;\;$      0.0     & 33.4 $\;\;\;\;\;\;$         0.0\\ 
    & &  $\sigma_{j_{1}}^{2}=0.25\;\;\;\sigma_{j_{2}}^{2}=0.01$&\\\cline{3-3}
1   &6.0 $\;\;\;\;\;\;$   0.6  &  12.6 $\;\;\;\;\;\;$    4.4   &   26.6  $\;\;\;\;\;\;$       38.0\\  
5   & 0.0 $\;\;\;\;\;\;$     0.0&      0.0 $\;\;\;\;\;\;$      0.0 &      5.2  $\;\;\;\;\;\;$     22.8\\ 
10  & 0.2   $\;\;\;\;\;\;$   0.0  &    0.0 $\;\;\;\;\;\;$      0.0 &        1.0 $\;\;\;\;\;\;$      19.6\\
20  & 0.2  $\;\;\;\;\;\;$    0.0 &     0.0 $\;\;\;\;\;\;$      0.0  &       0.0  $\;\;\;\;\;\;$       20.0\\
\hline
\end{tabular}
\end{table}

\subsection{Comparing scenarios with different degrees of cross-classification}
It is well known that omitting a crossed factor in a linear model yields a bias on the variance of the remaining factor \cite{fiel02,Meyers}. According to \cite{luokwok}, the direction and magnitude of the bias are related to the degree of cross-classification. In the evaluation of INLA we do not consider the omission of a factor, anyway it is worth to check whether the degree of cross-classification plays a role in the performance of the estimators.

Following \cite{luokwok}, we measure the degree of cross-classification by the number of receivers for a fixed number of feeders (see, for example, Table \ref{table:receiver1}). In particular, given 10 feeders, we consider three configurations, namely 10 receivers (complete cross-classification. i.e. without empty cells), 5 receivers (partial cross-classification with 50\% empty cells), and 2 receivers (partial cross-classification with 80\% empty cells, a situation close to a hierarchical structure). The three configurations have differing numbers of observations per cell to ensure a total sample size of 1000.
Table \ref{table:partial2} reports the results obtained under different degrees of cross-classification for a scenario where the random effects variances have magnitude equal to 0.25, in order to compare these results to those of Table \ref{table:complete0505}. The effect of the degree of cross-classification on the bias of the regression coefficients is conflicting, but remarkably modest. Therefore, in a situation with 1000 observations in a $10\times10$ matrix, the pattern of empty cells is practically uninfluential for the estimation of the regression coefficients: this is a noteworthy result since in many cross-classified datasets most of the cells are empty \cite{shi}.

\begin{table}
\caption{Relative bias for regression coefficients (relative bias of standard errors in parenthesis). Logistic model of equation (\ref{modello}) with $\sigma_{j_{1}}^{2}=\sigma_{j_{2}}^{2}=0.25$. Structures with different degrees of cross-classification (10 feeders and varying number of receivers), and varying number of observations per cell $n$.} \label{table:partial2}
\centering
\begin{tabular}{|cc|ccc|}
\hline
$n$ & Receivers& INLA Ga(0.001,0.001)& INLA Ga(0.5,0.003737)& MLLA\\
\hline
    & & & $\alpha$& \\\cline{4-4}
50  & 2& 0.230 (0.128)&  0.020 (0.023)&  0.220 (-0.051)\\
20  & 5& 0.010 (0.028)&   0.080 (0.244)&  0.000 (-0.116)\\
10  & 10&  -0.110 (0.102)& -0.030 (0.003)& -0.120 (-0.103)\\
    & & & $\beta_{1}$& \\\cline{4-4}
50  & 2&-0.020 (0.049)& -0.030 (-0.037)& -0.020 (0.054)\\
20  & 5&0.010 (0.016)&  0.070 (0.002)&  0.010 (0.021)\\
10  & 10&0.050 (-0.018)&  0.060 (0.065)&  0.040 (-0.012)\\
    & & & $\beta_{2}$& \\\cline{4-4}
50  & 2&-0.023 (-0.012)& 0.005 (-0.037)& -0.028 (-0.006)\\
20  & 5&0.017 (-0.020)& -0.003 (0.002)&  0.010 (-0.007)\\
10  & 10&0.025 (-0.006)&  0.003 (0.065)&  0.017 (0.008)\\
    & & & $\gamma_{1}$& \\\cline{4-4}
50  & 2&0.017 (0.115)&  0.032 (0.214)&  0.008 (-0.078)\\
20  & 5&-0.020 (0.069)&-0.050 (0.164)& -0.028 (-0.108)\\
10  & 10&0.008 (0.041)&  0.080 (-0.009)&  0.000 (-0.191)\\
    & & & $\gamma_{2}$& \\\cline{4-4}
50  & 2&0.000 (-0.056)& -0.005 (0.025)& -0.005 (-0.051)\\
20  & 5&0.025 (-0.011)&  0.060 (-0.015)&  0.020 (-0.006)\\
10  & 10&-0.008 (0.040)& 0.065  (-0.025)& -0.013 (-0.193)\\
\hline
\end{tabular}
\end{table}

\begin{figure}
\centerline{\includegraphics[scale=0.3]{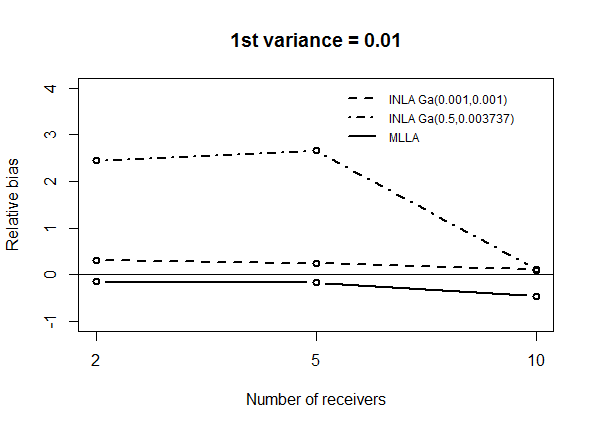}
            \includegraphics[scale=0.3]{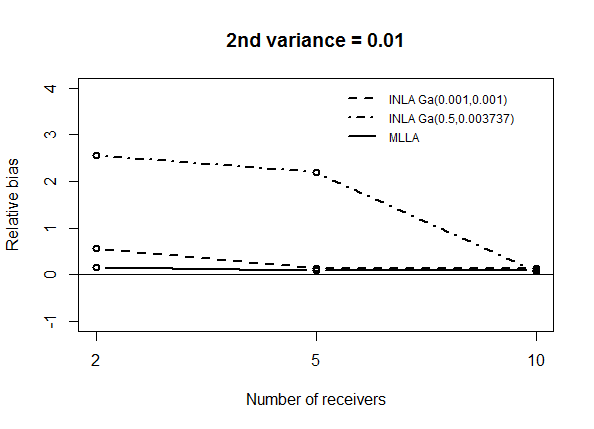} }
\centerline{\includegraphics[scale=0.3]{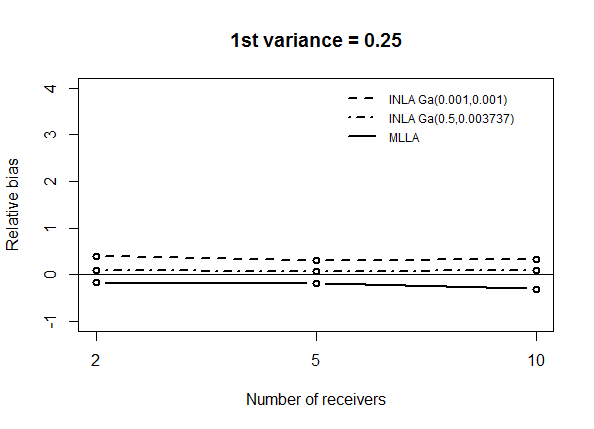}
            \includegraphics[scale=0.3]{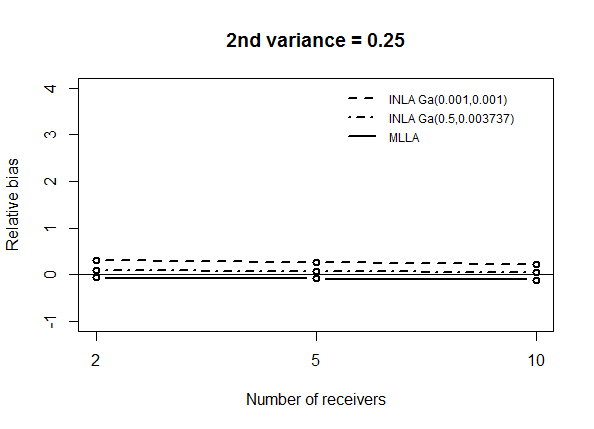}}
\centerline{\includegraphics[scale=0.3]{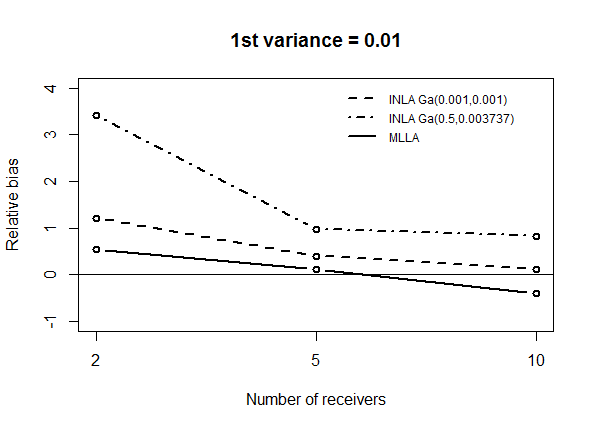}
            \includegraphics[scale=0.3]{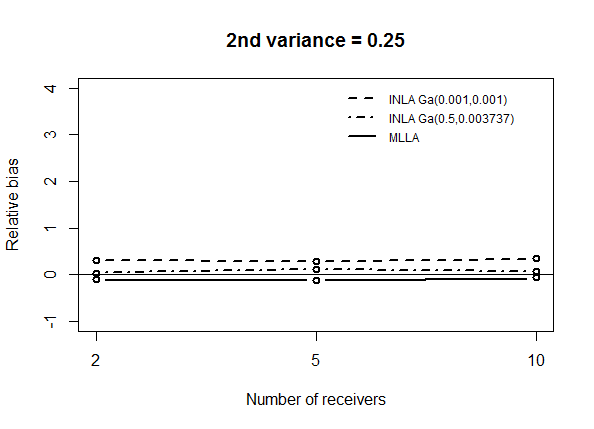}}
\centerline{\includegraphics[scale=0.3]{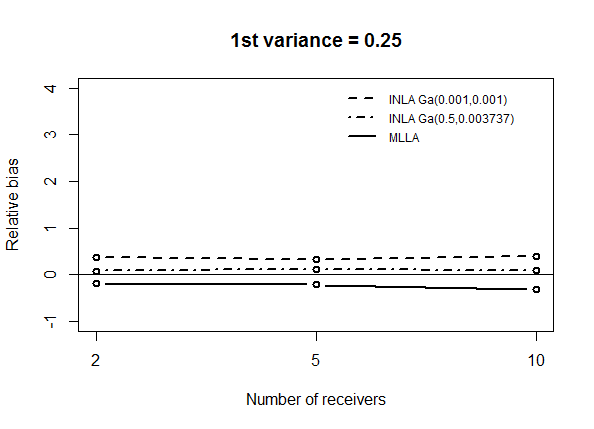}
            \includegraphics[scale=0.3]{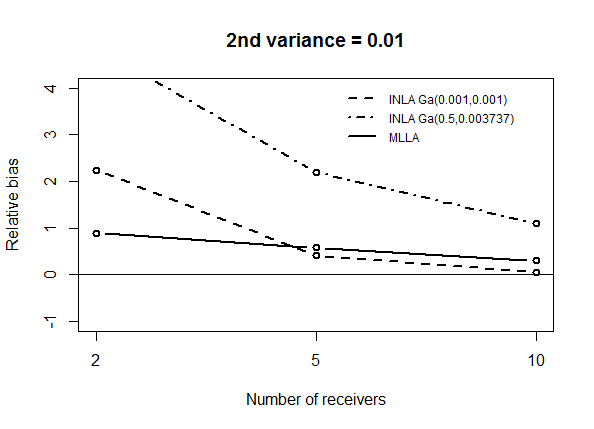} }
\caption{Relative bias for the variance components of the logistic model of equation (\ref{modello}). Structures with different degree of cross-classification (10 feeders and varying number of receivers). The cell sample size $n$ is set on the basis of the number of receivers to ensure a total sample size of 1000. Each pair of graphs corresponds to a combination of random effects variances ($\sigma_{j_{1}}^{2}, \sigma_{j_{2}}^{2}$): (0.01, 0.01), (0.25, 0.25), (0.01, 0.25), (0.25, 0.01).}
\label{figure:partially}
\end{figure}

The effect of the degree of cross-classification on the estimation of the random effects variances is shown in Figure \ref{figure:partially}. The effect is more sizable than for regression coefficients, especially if the true variance is tiny (0.01): in such instances, the performance of INLA improves as the cross-classification matrix becomes closer to completeness (10 receivers), especially for the prior Ga(0.5,0.003737). The effect on MLLA estimates is more conflicting because in some scenarios increasing the number of receivers implies a larger bias (even for variance equal to $0.25$), while in other scenarios the bias remains unchanged across the three degrees of cross-classification and in one case it slightly decreases (for the second variance in the scenario with $\sigma_{j_{1}}^{2}=0.25$ and $\sigma_{j_{2}}^{2}=0.01$). This contradictory behaviour can be explained by the high percentage of null estimates which undoubtedly affects the bias. It is worth to note that, for all the estimators, the degree of cross-classification plays a major role when the variances of the two factors are markedly different (0.01 and 0.25): in those cases, if the matrix is sparse (2 receivers) the estimation of the tiny variance is largely out of the target.

Table \ref{table:anomalie2} reports the percentages of extreme estimates in variance components for the considered estimators, showing that the degree of cross-classification has a negligible role. Note that in the considered scenarios, all having a total sample size of 1000, INLA produces almost no aberrant estimates, whereas MLLA yields many zero estimates when the variance component is tiny (0.01).

\begin{table}
\caption{Percentage of extreme estimates out of the 500 replicates. Structures with different degree of cross-classification (10 feeders and varying number of receivers), and varying number of observations per cell $n$.} \label{table:anomalie2}
\centering
\begin{tabular}{|cc|ccc|}
\hline
Receivers & $n$ &  INLA Ga(0.001,0.001)& INLA Ga(0.5,0.003737)& MLLA\\
\hline
    & &\% $\hat{\sigma}_{j_{1}}^{2}> 2$ \& \% $\hat{\sigma}_{j_{2}}^{2}> 2$ &\% $\hat{\sigma}_{j_{1}}^{2}> 2$ \& \% $\hat{\sigma}_{j_{2}}^{2}> 2$ &\% $\hat{\sigma}_{j_{1}}^{2}=0$ \& \% $\hat{\sigma}_{j_{2}}^{2}=0$ \\\cline{3-5}
    &  & & $\sigma_{j_{1}}^{2}=\sigma_{j_{2}}^{2}=0.01$& \\\cline{4-4}
 2  &50& 0.0   $\;\;\;\;\;\;$      0.0   &   0.0   $\;\;\;\;\;\;$       0.0    &  36.6 $\;\;\;\;\;\;$        35.0\\
 5  &20& 0.0   $\;\;\;\;\;\;$      0.0  &    0.0   $\;\;\;\;\;\;$       0.0    &  39.4  $\;\;\;\;\;\;$     30.6\\
10  &10&  0.0   $\;\;\;\;\;\;$      0.0  &    0.0   $\;\;\;\;\;\;$       0.0 &     44.6 $\;\;\;\;\;\;$      35.6\\
    &  & & $\sigma_{j_{1}}^{2}=\sigma_{j_{2}}^{2}=0.25$& \\\cline{4-4}
 2  &50& 0.2   $\;\;\;\;\;\;$   0.0   &   0.0   $\;\;\;\;\;\;$       0.0   &    1.4   $\;\;\;\;\;\;$     0.8\\
 5  &20&0.0   $\;\;\;\;\;\;$      0.0    &  0.0   $\;\;\;\;\;\;$       0.0     &  0.8  $\;\;\;\;\;\;$      0.6\\
10  &10&0.0   $\;\;\;\;\;\;$      0.0&      0.0   $\;\;\;\;\;\;$       0.0  &     0.8  $\;\;\;\;\;\;$        0.0\\
    &  & &$\sigma_{j_{1}}^{2}=0.01\;\;\;\sigma_{j_{2}}^{2}=0.25$& \\\cline{4-4}
 2  &50& 0.0   $\;\;\;\;\;\;$      0.0  &    0.0  $\;\;\;\;\;\;$     0.0    &  20.6   $\;\;\;\;\;\;$     0.4\\
 5  &20& 0.0   $\;\;\;\;\;\;$      0.0   &   0.0   $\;\;\;\;\;\;$    0.0 &     25.8   $\;\;\;\;\;\;$       0.0\\
10  &10&0.0   $\;\;\;\;\;\;$      0.0 &     0.0   $\;\;\;\;\;\;$    0.0   &   39.4   $\;\;\;\;\;\;$       0.0\\
    &  & &$\sigma_{j_{1}}^{2}=0.25\;\;\;\sigma_{j_{2}}^{2}=0.01$& \\\cline{4-4}
 2  &50&  0.0   $\;\;\;\;\;\;$      0.0  &    0.0   $\;\;\;\;\;\;$       0.0   &    0.8  $\;\;\;\;\;\;$     22.2\\
 5  &20&  0.0   $\;\;\;\;\;\;$      0.0  &    0.0   $\;\;\;\;\;\;$       0.0   &    0.8 $\;\;\;\;\;\;$        17.0\\
10  &10&0.2  $\;\;\;\;\;\;$    0.0  &    0.0   $\;\;\;\;\;\;$       0.0 &        1.0   $\;\;\;\;\;\;$       19.6\\
\hline
\end{tabular}
\end{table}
\clearpage

\subsection{Assessing the asymptotic behaviour}
In order to evaluate the asymptotic behaviour of the considered estimation methods, we increase the number of clusters per classification in a setting with complete cross-classification and constant cell sample size $n=10$.
For the regression coefficients, Table \ref{table:a5} shows that INLA and MLLA have similar, satisfactory performances: the relative biases are smaller than 5\% even with 20 clusters per classification and the differences between the two methods decline sharply as the number of clusters increases.

\begin{table}
\caption{Relative bias for regression coefficients (relative bias of standard errors in parenthesis). Logistic model of equation (\ref{modello}) with $\sigma_{j_{1}}^{2}=\sigma_{j_{2}}^{2}=0.25$. Complete cross-classification with varying number of clusters per classification $N_{1}=N_{2}$ and $n=10$ observations per cell.} \label{table:a5}
\centering
\begin{tabular}{|c|ccc|}
\hline
$N_{1}=N_{2}$ & INLA Ga(0.001,0.001)& INLA Ga(0.5,0.003737)& MLLA\\
\hline
  & & $\alpha$& \\\cline{3-3}
10&-0.110 (0.102)& -0.030 (0.003)& -0.120 (-0.103)\\
20& 0.080 (0.029)&  -0.070 (-0.022)&  0.080 (-0.046)\\
50&-0.020 (-0.013)& 0.140 (-0.021)& -0.020 (-0.041)\\
80&-0.010 (-0.009)& -0.040 (-0.067)& -0.010 (-0.020)\\
  & & $\beta_{1}$& \\\cline{3-3}
10&0.050 (-0.018)&  0.060 (-0.047)&  0.040 (-0.012)\\
20&0.030 (-0.018)&  0.020 (-0.009)&  0.030 (-0.017)\\
50&-0.010 (-0.039)& 0.010 (0.075)& -0.010 (-0.039)\\
80&0.000 (0.067)&  0.010 (-0.005)&  0.000 (0.067)\\
  & &$\beta_{2}$& \\\cline{3-3}
10&0.025 (-0.006)& 0.003 (0.065)&  0.017 (0.008)\\
20&-0.010 (0.088)& 0.003 (-0.011)& -0.013 (0.090)\\
50&0.005 (-0.016)&  -0.003 (-0.064)& 0.005 (-0.016)\\
80&0.000 (0.002)&  0.000 (0.011)&  0.000 (0.002)\\
  & & $\gamma_{1}$& \\\cline{3-3}
10&0.008 (0.041)&  0.080 (-0.009)&  0.000 (-0.191)\\
20&-0.018 (0.023)& 0.022 (-0.008)& -0.020 (-0.060)\\
50&0.020 (0.039)&  -0.050 (0.014)&  0.020 (0.012)\\
80&0.030 (-0.010)&  0.022 (-0.008)&  0.030 (-0.027)\\
  & & $\gamma_{2}$& \\\cline{3-3}
10&-0.008 (0.040)& 0.065 (-0.025)& -0.013 (-0.193)\\
20&0.005 (-0.018)& -0.025 (-0.026)&  0.003 (-0.018)\\
50&0.003 (-0.027)& 0.000 (-0.045)&  0.003 (0.002)\\
80&0.000 (-0.036)& 0.003 (-0.010)&  0.000 (-0.036)\\
\hline
\end{tabular}
\end{table}

\begin{figure}
\centerline{\includegraphics[scale=0.3]{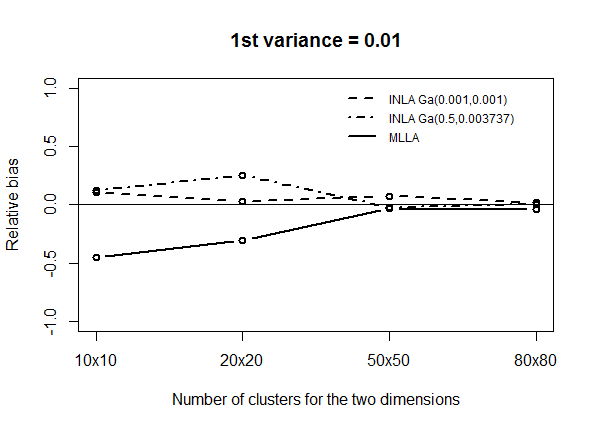} 
            \includegraphics[scale=0.3]{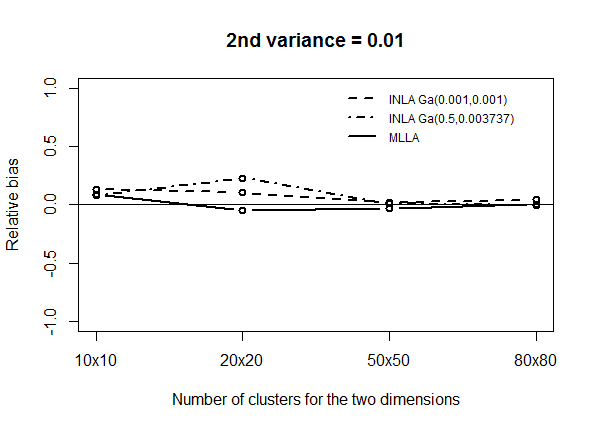} }
\centerline{\includegraphics[scale=0.3]{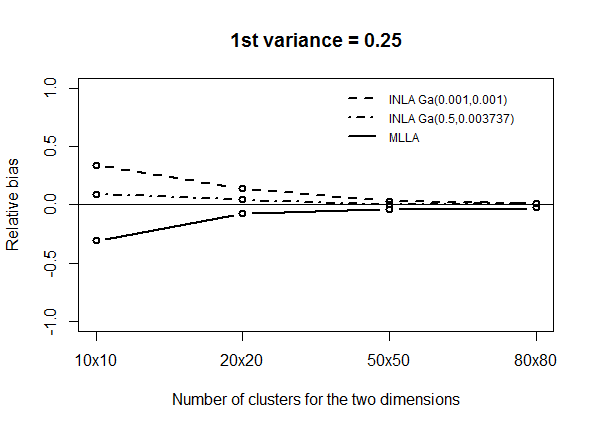} 
            \includegraphics[scale=0.3]{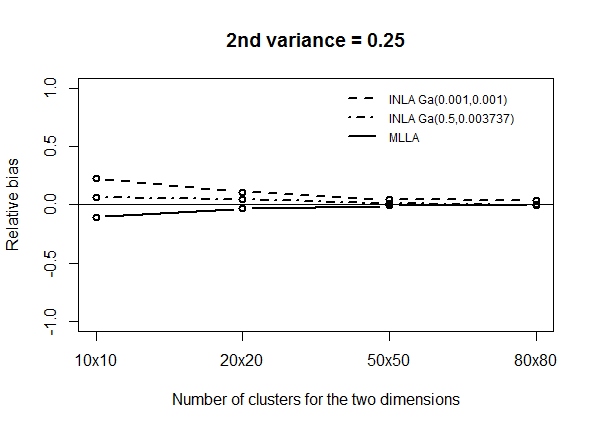} }
\centerline{\includegraphics[scale=0.3]{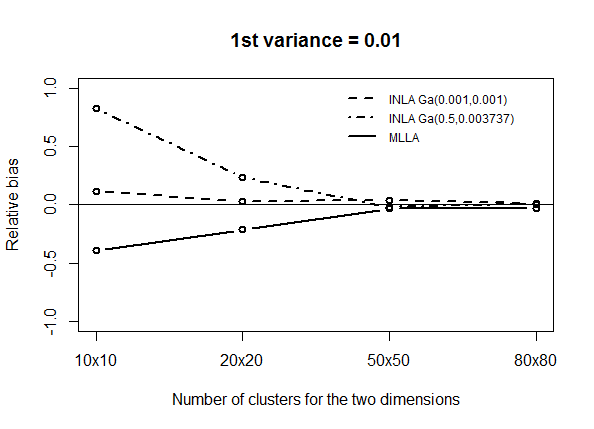} 
            \includegraphics[scale=0.3]{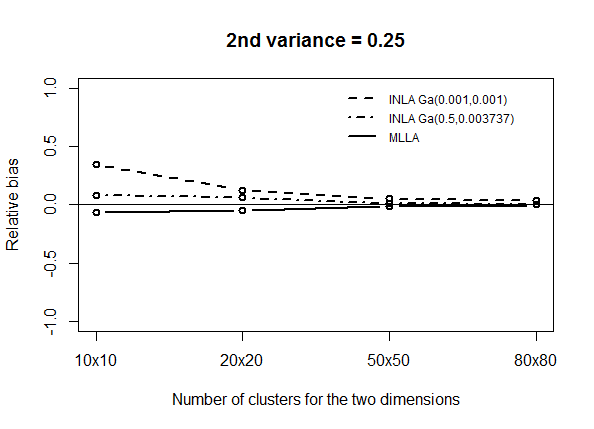} }
\centerline{\includegraphics[scale=0.3]{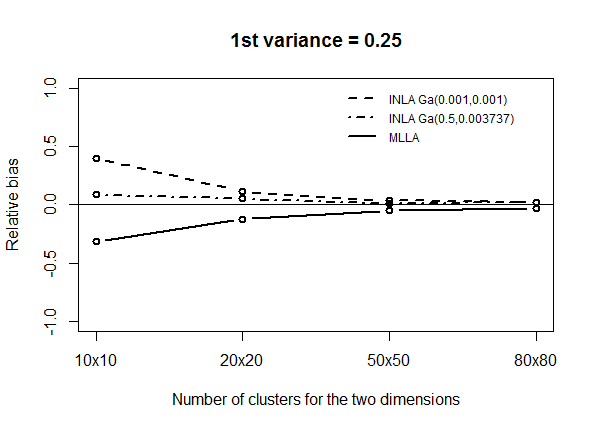} 
            \includegraphics[scale=0.3]{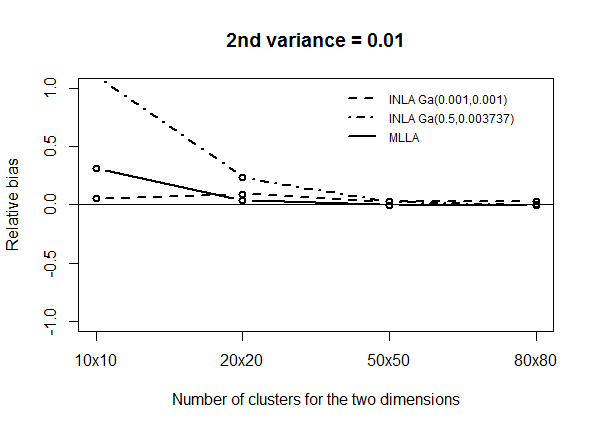} }
\caption{Relative bias for the variance components of the logistic model of equation (\ref{modello}). Complete cross-classification with varying numbers of clusters per dimension $N_{1} \times N_{2}$ and $n=10$ observations per cell. Each pair of graphs corresponds to a combination of random effects variances ($\sigma_{j_{1}}^{2}, \sigma_{j_{2}}^{2}$): (0.01, 0.01), (0.25, 0.25), (0.01, 0.25), (0.25, 0.01).}
\label{figure:asymptotics1}
\end{figure}

Also for the variance components (Figure \ref{figure:asymptotics1}) the results become similar as the number of clusters increases, though the requirement for a satisfactory performance is higher (50 clusters per classification).
As for the priors, INLA Ga(0.5,0.003737) has the best performance among the three estimators when $N_{1}=N_{2}\geq 50$ regardless of the magnitude of variances, and when $N_{1}=N_{2}\leq 20$ if at least one variance is $0.25$. On the other hand, when the variance components are tiny and the dimensions of the cross-classification matrix are small (less than $20$ clusters), INLA Ga(0.001,0.001) yields the smallest bias.
Although the differences between Bayesian and frequentist methods vanish asymptotically, in the considered settings MLLA globally has the worst performance, particularly for big variances ($0.25$). Moreover, the direction of its bias is hardly predictable when $N_{1}=N_{2}\leq 20$ and at least a variance is $0.01$, since it is negative for the first variance and positive for the second one.

The percentages of extreme estimates for the considered methods are reported in Table \ref{table:anomalie3}. It is worth to note that MLLA needs more than 4000 observations to yield less than 20\% of zero estimates (when the magnitude of the variances is $0.01$), for INLA the issue of aberrant estimates vanishes with 500 observations regardless of the size of the variance components.

\begin{table}
\caption{Percentage of extreme estimates out of the 500 replicates. Complete cross-classification with varying number of clusters per classification $N_{1}=N_{2}$ and $n=10$ observations per cell.} \label{table:anomalie3}
\centering
\begin{tabular}{|c|ccc|}
\hline
$N_{1}=N_{2}$ & INLA Ga(0.001,0.001)& INLA Ga(0.5,0.003737)& MLLA\\
\hline
 &\% $\hat{\sigma}_{j_{1}}^{2}> 2$ \& \% $\hat{\sigma}_{j_{2}}^{2}> 2$ &\% $\hat{\sigma}_{j_{1}}^{2}> 2$ \& \% $\hat{\sigma}_{j_{2}}^{2}> 2$ &\% $\hat{\sigma}_{j_{1}}^{2}=0$ \& \% $\hat{\sigma}_{j_{2}}^{2}=0$ \\
\cline{2-4}
  &                             & $\sigma_{j_{1}}^{2}=\sigma_{j_{2}}^{2}=0.01$& \\\cline{3-3}
10& 0.0  $\;\;\;\;\;\;$    0.0  &    0.0 $\;\;\;\;\;\;$      0.0 &     44.6   $\;\;\;\;\;\;$    35.6\\
20& 0.0  $\;\;\;\;\;\;$    0.0  &     0.0  $\;\;\;\;\;\;$     0.0  &    19.8   $\;\;\;\;\;\;$    12.4\\
50& 0.0  $\;\;\;\;\;\;$    0.0  &    0.0  $\;\;\;\;\;\;$     0.0   &      0.0    $\;\;\;\;\;\;$      0.0\\
80& 0.0  $\;\;\;\;\;\;$    0.0  &     0.0  $\;\;\;\;\;\;$     0.0   &      0.0     $\;\;\;\;\;\;$     0.0\\
  &                             & $\sigma_{j_{1}}^{2}=\sigma_{j_{2}}^{2}=0.25$& \\\cline{3-3}
10& 0.0  $\;\;\;\;\;\;$    0.0  &      0.0  $\;\;\;\;\;\;$     0.0  &     0.8     $\;\;\;\;\;\;$     0.0\\
20& 0.0  $\;\;\;\;\;\;$    0.0  &     0.0   $\;\;\;\;\;\;$    0.0  &       0.0    $\;\;\;\;\;\;$      0.0\\
50& 0.0   $\;\;\;\;\;\;$   0.0  &    0.0  $\;\;\;\;\;\;$     0.0  &       0.0    $\;\;\;\;\;\;$      0.0\\
80& 0.0  $\;\;\;\;\;\;$    0.0  &    0.0   $\;\;\;\;\;\;$    0.0  &       0.0   $\;\;\;\;\;\;$       0.0\\
  &                             &$\sigma_{j_{1}}^{2}=0.01\;\;\;\sigma_{j_{2}}^{2}=0.25$& \\\cline{3-3}
10& 0.0   $\;\;\;\;\;\;$   0.0  &     0.0   $\;\;\;\;\;\;$    0.0   &   39.4     $\;\;\;\;\;\;$     0.0\\
20& 0.0  $\;\;\;\;\;\;$    0.0  &    0.0   $\;\;\;\;\;\;$    0.0 &     17.2     $\;\;\;\;\;\;$     0.0\\
50& 0.0  $\;\;\;\;\;\;$    0.0  &    0.0  $\;\;\;\;\;\;$     0.0  &       0.0     $\;\;\;\;\;\;$     0.0\\
80& 0.0   $\;\;\;\;\;\;$   0.0  &    0.0 $\;\;\;\;\;\;$      0.0  &       0.0    $\;\;\;\;\;\;$     0.0\\
  &                             & $\sigma_{j_{1}}^{2}=0.25\;\;\;\sigma_{j_{2}}^{2}=0.01$& \\\cline{3-3}
10& 0.2  $\;\;\;\;\;\;$    0.0  &    0.0   $\;\;\;\;\;\;$    0.0 &        1.0    $\;\;\;\;\;\;$   19.6\\
20& 0.0   $\;\;\;\;\;\;$   0.0  &    0.0   $\;\;\;\;\;\;$    0.0  &       0.0    $\;\;\;\;\;\;$     12.0\\
50& 0.0  $\;\;\;\;\;\;$    0.0  &    0.0  $\;\;\;\;\;\;$     0.0  &       0.0     $\;\;\;\;\;\;$     0.0\\
80& 0.0   $\;\;\;\;\;\;$   0.0  &    0.0  $\;\;\;\;\;\;$     0.0  &       0.0     $\;\;\;\;\;\;$     0.0\\
\hline
\end{tabular}
\end{table}

\section{The Salamander mating data}
\label{salam}
In this section we summarize the results obtained by applying INLA to the famous salamander mating data, which have become a standard test for estimation methods of cross-classified logistic models. The salamander mating data, presented for the first time by \cite{salamander}, were collected in 1986 by S. Arnold and P. Verell of the University of Chicago, Department of Ecology and Evolution, through three experiments on forty mountain dusky salamanders belonging to two different populations. The two populations, called Rough Butt and Whiteside from the names of the locations where they lived, were geographically isolated from one another, thus the aim of the three experiments was to investigate the extent to which Rough Butt and Whiteside would interbreed. In each experiment 40 salamanders were involved: they were divided in two groups each composed by five Rough Butt males, five Rough Butt females, five Whiteside males and five Whiteside females. Each salamander was paired with six partners, three belonging to the same population and three from the other, then across the three experiments 360 pairs were formed.\\
We consider model A of \cite{zeger}, which is a two-level random intercept logistic cross-classified model similar to the one defined by equation (\ref{modello}). Specifically, the model for salamander mating has two covariates at level 2 and an interaction term:
\begin{equation} \label{modello salamandre}
logit(\pi_{i(j_{1}j_{2})})=\beta_{0}+\beta_{1}x_{1j_{1}}+\beta_{2}x_{2j_{2}}+\beta_{3}x_{1j_{1}}x_{2j_{2}}+
    u_{j_{1}}+u_{j_{2}}
\end{equation}
\[u_{j_{1}}\sim N(0,\sigma^{2}_{u_{j_{1}}})\;\;\; u_{j_{2}}\sim N(0,\sigma^{2}_{u_{j_{2}}})\]\\
where $\pi_{i(j_{1}j_{2})}$ is the probability of a successful mating between male $j_{1}$ (crossing factor 1) and female $j_{2}$ (crossing factor 2). The binary covariates $x_{1j_{1}}$ and $x_{2j_{2}}$ take the value 1 if the salamander is a Whiteside male or Whiteside female, respectively. Each factor of classification is composed by 60 clusters, within each cluster there are 6 level 1 units (i.e. male-female pairs), within each cell or pair of clusters belonging to the two factors there is a single level 1 unit. 
The data structure is sketched in Table \ref{table:salamander}, with 60 rows for the females, 60 columns for the males, and 6 blocks representing groups of salamanders across the three experiments. Therefore, the data are partially cross-classified with no replications within cells (i.e. the cell sample size is one) and 90\% of empty cells. This structure is quite different from the structures considered in our simulation study, thus the findings of Section \ref{results} do not necessarily carry over.

\begin{table}
\caption{The Salamander mating data structure: columns represent factor 1 (male), rows represent factor 2 (female).} \label{table:salamander}
\centering\resizebox{\textwidth}{!}{  
\begin{tabular}{|c|cccccccccc
                   cccccccccc
                   cccccccccc
                   cccccccccc
                   cccccccccc
                   cccccccccc|}
\hline
&1&2&3&4&5&6&7&8&9&10&11&12&13&14&15&16&17&18&19&20&21&22&23&24&25&26&27&28&29&30
&31&32&33&34&35&36&37&38&39&40&41&42&43&44&45&46&47&48&49&50&51&52&53&54&55&56
&57&58&59&60\\\hline
1 &x&&&x&x&x&&&x&x&&&&&&&&&&&&&&&&&&&&&&&&&&&&&&&&&&&&&&&&&&&&&&&&&&\\
2 &x&&x&&x&&x&x&&x&&&&&&&&&&&&&&&&&&&&&&&&&&&&&&&&&&&&&&&&&&&&&&&&&&\\
3 &x&x&x&&&x&&x&x&&&&&&&&&&&&&&&&&&&&&&&&&&&&&&&&&&&&&&&&&&&&&&&&&&&\\
4 &&x&&x&x&&x&&x&x&&&&&&&&&&&&&&&&&&&&&&&&&&&&&&&&&&&&&&&&&&&&&&&&&&\\
5 &&x&x&x&&x&x&x&&&&&&&&&&&&&&&&&&&&&&&&&&&&&&&&&&&&&&&&&&&&&&&&&&&&\\
6 &x&x&&x&&&x&x&&x&&&&&&&&&&&&&&&&&&&&&&&&&&&&&&&&&&&&&&&&&&&&&&&&&&\\
7 &x&x&&&x&x&&&x&x&&&&&&&&&&&&&&&&&&&&&&&&&&&&&&&&&&&&&&&&&&&&&&&&&&\\
8 &&&x&x&x&x&x&&&x&&&&&&&&&&&&&&&&&&&&&&&&&&&&&&&&&&&&&&&&&&&&&&&&&&\\
9 &x&&x&x&&x&&x&x&&&&&&&&&&&&&&&&&&&&&&&&&&&&&&&&&&&&&&&&&&&&&&&&&&&\\
10&&x&x&&x&&x&x&x&&&&&&&&&&&&&&&&&&&&&&&&&&&&&&&&&&&&&&&&&&&&&&&&&&&\\
11&&&&&&&&&& &&x&x&x&&x&&&x&x&&&&&&&&&&&&&&&&&&&&&&&&&&&&&&&&&&&&&&&&\\
12&&&&&&&&&& &x&&x&x&&&x&x&x&&&&&&&&&&&&&&&&&&&&&&&&&&&&&&&&&&&&&&&&&\\
13&&&&&&&&&& &x&&&x&x&x&x&&&x&&&&&&&&&&&&&&&&&&&&&&&&&&&&&&&&&&&&&&&&\\
14&&&&&&&&&& &x&x&&&x&&&x&x&x&&&&&&&&&&&&&&&&&&&&&&&&&&&&&&&&&&&&&&&&\\
15&&&&&&&&&& &&x&x&&x&x&x&x&&&&&&&&&&&&&&&&&&&&&&&&&&&&&&&&&&&&&&&&&&\\
16&&&&&&&&&& &&x&&x&x&&&x&x&x&&&&&&&&&&&&&&&&&&&&&&&&&&&&&&&&&&&&&&&&\\
17&&&&&&&&&& &x&x&&x&&x&x&&&x&&&&&&&&&&&&&&&&&&&&&&&&&&&&&&&&&&&&&&&&\\
18&&&&&&&&&& &x&x&x&&&x&x&&x&&&&&&&&&&&&&&&&&&&&&&&&&&&&&&&&&&&&&&&&&\\
19&&&&&&&&&& &&&x&x&x&&&x&x&x&&&&&&&&&&&&&&&&&&&&&&&&&&&&&&&&&&&&&&&&\\
20&&&&&&&&&& &x&&x&&x&x&x&x&&&&&&&&&&&&&&&&&&&&&&&&&&&&&&&&&&&&&&&&&&\\
21&&&&&&&&&&&&&&&&&&&& &x&&&x&x&x&&&x&x&&&&&&&&&&&&&&&&&&&&&&&&&&&&&&\\
22&&&&&&&&&&&&&&&&&&&& &x&&x&&x&&x&x&&x&&&&&&&&&&&&&&&&&&&&&&&&&&&&&&\\
23&&&&&&&&&&&&&&&&&&&& &x&x&x&&&x&&x&x&&&&&&&&&&&&&&&&&&&&&&&&&&&&&&&\\
24&&&&&&&&&&&&&&&&&&&& &&x&&x&x&&x&&x&x&&&&&&&&&&&&&&&&&&&&&&&&&&&&&&\\
25&&&&&&&&&&&&&&&&&&&& &&x&x&x&&x&x&x&&&&&&&&&&&&&&&&&&&&&&&&&&&&&&&&\\
26&&&&&&&&&&&&&&&&&&&& &x&x&&x&&&x&x&&x&&&&&&&&&&&&&&&&&&&&&&&&&&&&&&\\
27&&&&&&&&&&&&&&&&&&&& &x&x&&&x&x&&&x&x&&&&&&&&&&&&&&&&&&&&&&&&&&&&&&\\
28&&&&&&&&&&&&&&&&&&&& &&&x&x&x&x&x&&&x&&&&&&&&&&&&&&&&&&&&&&&&&&&&&&\\
29&&&&&&&&&&&&&&&&&&&& &x&&x&x&&x&&x&x&&&&&&&&&&&&&&&&&&&&&&&&&&&&&&&\\
30&&&&&&&&&&&&&&&&&&&& &&x&x&&x&&x&x&x&&&&&&&&&&&&&&&&&&&&&&&&&&&&&&&\\
31&&&&&&&&&&&&&&&&&&&&&&&&&&&&&& &&x&x&x&&x&&&x&x&&&&&&&&&&&&&&&&&&&&\\
32&&&&&&&&&&&&&&&&&&&&&&&&&&&&&& &x&&x&x&&&x&x&x&&&&&&&&&&&&&&&&&&&&&\\
33&&&&&&&&&&&&&&&&&&&&&&&&&&&&&& &x&&&x&x&x&x&&&x&&&&&&&&&&&&&&&&&&&&\\
34&&&&&&&&&&&&&&&&&&&&&&&&&&&&&& &x&x&&&x&&&x&x&x&&&&&&&&&&&&&&&&&&&&\\
35&&&&&&&&&&&&&&&&&&&&&&&&&&&&&& &&x&x&&x&x&x&x&&&&&&&&&&&&&&&&&&&&&&\\
36&&&&&&&&&&&&&&&&&&&&&&&&&&&&&& &&x&&x&x&&&x&x&x&&&&&&&&&&&&&&&&&&&&\\
37&&&&&&&&&&&&&&&&&&&&&&&&&&&&&& &x&x&&x&&x&x&&&x&&&&&&&&&&&&&&&&&&&&\\
38&&&&&&&&&&&&&&&&&&&&&&&&&&&&&& &x&x&x&&&x&x&&x&&&&&&&&&&&&&&&&&&&&&\\
39&&&&&&&&&&&&&&&&&&&&&&&&&&&&&& &&&x&x&x&&&x&x&x&&&&&&&&&&&&&&&&&&&&\\
40&&&&&&&&&&&&&&&&&&&&&&&&&&&&&& &x&&x&&x&x&x&x&&&&&&&&&&&&&&&&&&&&&&\\
41&&&&&&&&&&&&&&&&&&&&&&&&&&&&&&&&&&&&&&&& &x&&&x&x&x&&&x&x&&&&&&&&&&\\
42&&&&&&&&&&&&&&&&&&&&&&&&&&&&&&&&&&&&&&&& &x&&x&&x&&x&x&&x&&&&&&&&&&\\
43&&&&&&&&&&&&&&&&&&&&&&&&&&&&&&&&&&&&&&&& &x&x&x&&&x&&x&x&&&&&&&&&&&\\
44&&&&&&&&&&&&&&&&&&&&&&&&&&&&&&&&&&&&&&&& &&x&&x&x&&x&&x&x&&&&&&&&&&\\
45&&&&&&&&&&&&&&&&&&&&&&&&&&&&&&&&&&&&&&&& &&x&x&x&&x&x&x&&&&&&&&&&&&\\
46&&&&&&&&&&&&&&&&&&&&&&&&&&&&&&&&&&&&&&&& &x&x&&x&&&x&x&&x&&&&&&&&&&\\
47&&&&&&&&&&&&&&&&&&&&&&&&&&&&&&&&&&&&&&&& &x&x&&&x&x&&&x&x&&&&&&&&&&\\
48&&&&&&&&&&&&&&&&&&&&&&&&&&&&&&&&&&&&&&&& &&&x&x&x&x&x&&&x&&&&&&&&&&\\
49&&&&&&&&&&&&&&&&&&&&&&&&&&&&&&&&&&&&&&&& &x&&x&x&&x&&x&x&&&&&&&&&&&\\
50&&&&&&&&&&&&&&&&&&&&&&&&&&&&&&&&&&&&&&&& &&x&x&&x&&x&x&x&&&&&&&&&&&\\
51&&&&&&&&&&&&&&&&&&&&&&&&&&&&&&&&&&&&&&&&&&&&&&&&&& &&x&x&x&&x&&&x&x\\
52&&&&&&&&&&&&&&&&&&&&&&&&&&&&&&&&&&&&&&&&&&&&&&&&&& &x&&x&x&&&x&x&x&\\
53&&&&&&&&&&&&&&&&&&&&&&&&&&&&&&&&&&&&&&&&&&&&&&&&&& &x&&&x&x&x&x&&&x\\
54&&&&&&&&&&&&&&&&&&&&&&&&&&&&&&&&&&&&&&&&&&&&&&&&&& &x&x&&&x&&&x&x&x\\
55&&&&&&&&&&&&&&&&&&&&&&&&&&&&&&&&&&&&&&&&&&&&&&&&&& &&x&x&&x&x&x&x&&\\
56&&&&&&&&&&&&&&&&&&&&&&&&&&&&&&&&&&&&&&&&&&&&&&&&&& &&x&&x&x&&&x&x&x\\
57&&&&&&&&&&&&&&&&&&&&&&&&&&&&&&&&&&&&&&&&&&&&&&&&&& &x&x&&x&&x&x&&&x\\
58&&&&&&&&&&&&&&&&&&&&&&&&&&&&&&&&&&&&&&&&&&&&&&&&&& &x&x&x&&&x&x&&x&\\
59&&&&&&&&&&&&&&&&&&&&&&&&&&&&&&&&&&&&&&&&&&&&&&&&&& &&&x&x&x&&&x&x&x\\
60&&&&&&&&&&&&&&&&&&&&&&&&&&&&&&&&&&&&&&&&&&&&&&&&&& &x&&x&&x&x&x&x&&\\
\hline
\end{tabular}}
\end{table}

Table \ref{table:ris salam} reports the point estimates of the parameters of model (\ref{modello salamandre}) for several estimation methods. We consider MLLA (as in our simulation study), MCEM (taken by \cite{cho,statamanu} as the gold-standard), AIP with AGQ (the most accurate algorithm in \cite{cho}), as well as the Bayesian MCMC estimates of \cite{zeger} obtained with uniform priors for both variance components (results with other priors are reported in Table 3 of \cite{cho}). 

\begin{table}
\caption{Results for salamander data. Standard errors in parenthesis.} \label{table:ris salam}
\centering
\resizebox{0.95\textwidth}{!}{
\begin{tabular}{|c|cccccc|}
\hline
&MCEM$^{\rm a}$ & AIP with AGQ$^{\rm a}$ & MLLA$^{\rm a}$& MCMC$^{\rm b}$ & INLA Ga(0.001,0.001)& INLA Ga(0.5,0.003737) \\
\hline
$\beta_{0}$ &1.02 & 1.02(0.41)  &1.00(0.37) &1.03(0.43) &1.01(0.40) &0.98(0.38) \\
$\beta_{1}$ &-0.69&-0.70(0.48)  &-0.70(0.44)&-0.69(0.50)&-0.69(0.45)&-0.67(0.43) \\
$\beta_{2}$ &-2.96& -2.96(0.58) &-2.91(0.50)&-3.01(0.60)&-2.94(0.55)&-2.84(0.54) \\
$\beta_{3}$ &3.63 &3.64(0.65)   &3.59(0.54) & 3.74(0.68)&3.61(0.60) &3.50(0.59) \\
$\sigma_{1}$&1.12 &1.11         &1.03       & 1.17      &1.10       &1.02       \\
$\sigma_{2}$&1.18 &1.17         &1.08       &1.22       &1.17       &1.10       \\
\hline
\end{tabular}}
\small{\begin{flushleft}$^{\rm a}$ From \cite{cho} Tables 1 and 2.
\end{flushleft}}
\small{\begin{flushleft}$^{\rm b}$ From \cite{zeger} Table 3 (uniform prior; posterior medians; SE is the range of the $90\%$ CI divided by 3.3)\end{flushleft}}
\end{table}

Table \ref{table:ris salam} shows that, taking MCEM as the benchmark, INLA with prior Ga(0.001,0.001) has a good performance, similar to that of AIP with AGQ, and better than MLLA and MCMC with uniform prior. On the other hand, INLA with prior INLA Ga(0.5,0.003737) has a less satisfactory performance, similarly to the results reported in the Supplementary Material of \cite{fong}. Overall, INLA is a valuable method also in the peculiar framework of salamander data, confirming the encouraging findings of simulation studies (Section \ref{results} for cross-classified models and \cite{grilli} for nested random effects). In addition to accuracy, INLA is considerably faster than MCEM, AIP and MCMC, indeed computation with the salamander data requires only a few seconds.

\section{Conclusions}
\label{conclusion}
We investigated the performance of INLA for fitting two-level random intercept logistic models with crossed random effects. The investigation exploited a detailed simulation study, entailing a comparison with the standard maximum likelihood method (MLLA), and an application to the classical salamander data, entailing a comparison with several competing methods (MCEM, AIP and MCMC). 

In the simulation study we devoted special attention to scenarios with a small number of clusters, varying degrees of cross-classification, small magnitudes of random effects variances and different prior specifications. Both INLA and MLLA give biases less than 5\% for the fixed effects even in scenarios with a small number of clusters and it turns out that 5 units per cell are enough to have good estimates. However, the cell sample size should be greater when random effects variances are larger than 0.25 and/or the cross-classification matrix of the data is sparse.
The simulation study showed that INLA and MLLA sometimes fail in estimating the variance components, though in a different way. Specifically, MLLA can yield zero estimates, particularly when a variance component is close to zero and the sample size is lower than 4000; on the other hand, INLA never yields zero estimates, though it occasionally provides aberrant estimates of the variance components in scenarios with few clusters ($10\times10$) and one observation per cell. This problem, though relevant only in few scenarios, should be further investigated in order to prevent it.
Moreover, in the three considered sets of scenarios (small number of clusters, different degrees of cross-classification, increasing number of clusters per classification), INLA is generally more accurate than MLLA: when the variance is $0.25$, INLA Ga(0.5,0.003737) has the best performance regardless of the data structure; when the variance is $0.01$, INLA Ga(0.001,0.001) performs better than MLLA (unless the cross-classification matrix is $10\times 10$ and the number of observations per cell is lower than 10). 
Moreover, INLA tends to overestimate the variance components, whereas MLLA showed the well-known downward bias \cite{cho,joe}. 
The degree of cross-classification has a little role on the performance of the estimators, even if in settings with small variances INLA yields better results when the data structure is closer to complete cross-classification.

The application to the salamander data showed that INLA is competitive with respect to the most efficient algorithms for cross-classified random effects (MCEM, AIP with AGQ, MCMC) since it has similar accuracy, but lower computational time.

In general, INLA has the advantage of directly approximating the posterior distribution, thus avoiding the subtle issue of assessing the convergence as in MCMC and AIP. A difficulty in the application of INLA, common to all Bayesian methods, is the specification of a suitable prior distribution for the random effects variances, especially in settings where such specification is able to substantially affect the results. Nonetheless, this difficulty is alleviated with INLA since its computational speed enables a thorough sensitivity analysis (\cite{roos1,roos2}).

To conclude, INLA is an effective method for fitting logistic models with crossed random effects. It is preferable to MLLA in terms of accuracy and to MCMC in terms of speed and simplicity of implementation. In settings with limited sample sizes all the methods have difficulties, which may hopefully be attenuated by improvements in the algorithms and in the specification of the prior distribution, see \cite{improveInla} and \cite{metaInla} for recent developments in INLA.

\newpage

\large{\textbf{\center Fitting logistic multilevel models with crossed random effects via Bayesian Integrated Nested Laplace Approximations: a simulation study.}}
\large{\textbf{\center Supplementary material.}}

\setcounter{page}{1}
\setcounter{section}{0}
\setcounter{table}{0}
\setcounter{figure}{0}
\vspace{1cm}

\section{Relative bias for the regression coefficients - scenarios with variance components equal to (0.01,0.01), (0.25,0.01) and (0.01,0.25)}
In this section we report the relative bias for the regression coefficients in scenarios where only the results on the variance components have been shown in the paper, namely the scenarios with variance components ($\sigma_{j_{1}}^{2}, \sigma_{j_{2}}^{2}$) equal to (0.01,0.01), (0.25,0.01) and (0.01,0.25). To facilitate comparisons, the subsequent subsections are structured as in the paper.

\subsection{Scenarios with few clusters}
Tables \ref{table:1}, \ref{table:2} and \ref{table:3} show the relative bias for the fixed effects of model (1), with variance components equal to, respectively, (0.01,0.01), (0.01,0.25) and (0.25,0.01), obtained in scenarios with complete cross-classification, $10$ clusters per classification and increasing number of observations per cell $n$.

\begin{table}
\caption{Relative bias for regression coefficients (relative bias of standard errors in parenthesis). Logistic model of equation (1) with $\sigma_{j_{1}}^{2}=\sigma_{j_{2}}^{2}=0.01$. Complete cross-classification with $N_{1}=N_{2}=10$ and varying number of observations per cell $n$.}\label{table:1}
\centering
\begin{tabular}{|c|ccc|}
\hline
$n$ & INLA Ga(0.001,0.001)& INLA Ga(0.5,0.003737)& MLLA\\
\hline
& & $\alpha$&\\\cline{3-3}
1&  -0.070 (0.134)&  -0.283 (-0.004)& -0.110 (0.028)\\
5& 0.020 (0.256)&  0.016 (0.147)&  0.000 (0.088)\\
10& 0.010 (0.161)&  -0.030 (0.142)&  0.000 (-0.029)\\ 
20& 0.010 (0.166)&  0.010 (0.097)&  0.000 (-0.050)\\ 
& & $\beta_{1}$&\\\cline{3-3}
1&0.366 (-0.132)& 0.239 (-0.087)&  0.280 (-0.071)\\ 
5&0.020 (-0.038)& 0.017 (-0.037)&  0.010 (-0.024)\\ 
10&0.000 (-0.026)& -0.030 (0.010)& -0.010 (-0.018)\\ 
20&0.030 (-0.025)& 0.030 (-0.020)&  0.030 (-0.019)\\ 
& & $\beta_{2}$&\\\cline{3-3}
1&0.154 (-0.095)& 0.112 (-0.069)& 0.062 (-0.039)\\
5&0.055 (0.023)& 0.049 (0.026)& 0.037 (0.039)\\
10&0.008 (0.073)& 0.027 (-0.008)& 0.000 (0.084)\\ 
20&0.005 (-0.050)& 0.004 (-0.046)& 0.000 (-0.045)\\ 
& & $\gamma_{1}$&\\\cline{3-3}
1&0.220 (0.103)&   0.179 (-0.041)&  0.120 (0.008)\\  
5&0.040 (0.200)&  0.032 (0.098)&  0.020 (0.034)\\ 
10& 0.020 (0.138)&  0.005 ( 0.120)&  0.013 (-0.051)\\
20&-0.005 (0.208)& -0.007 (0.134)& -0.010 (-0.023)\\     
& & $\gamma_{2}$&\\\cline{3-3}
1&0.210 (0.111)&  0.167 (0.007)&  0.110 (0.057)\\
5&0.013 (-0.008)&  0.006  (-0.037)&  0.000 (-0.044)\\
10&0.020 (0.152)&   -0.010 (0.124)&  0.010 (-0.045)\\
20&-0.003 (0.045)&  -0.004 (0.034)& -0.005 (-0.007)\\ 
\hline
\end{tabular}
\end{table}

\begin{table}
\caption{Relative bias for regression coefficients (relative bias of standard errors in parenthesis). Logistic model of equation (1) with $\sigma_{j_{1}}^{2}=0.01$ and $\sigma_{j_{2}}^{2}=0.25$. Complete cross-classification with $N_{1}=N_{2}=10$ and varying number of observations per cell $n$.}\label{table:2}
\centering
\begin{tabular}{|c|ccc|}
\hline
$n$ & INLA Ga(0.001,0.001)& INLA Ga(0.5,0.003737)& MLLA\\
\hline
& & $\alpha$& \\\cline{3-3}
1&  0.251 (0.073)&  0.061 (-0.099)&  0.078 (-0.029)\\
5& -0.011 (0.164)& 0.020  (0.033)& -0.045 (0.010)\\
10& 0.000 (0.176)&  0.170  (0.042)& -0.010 (0.018)\\
20& 0.010 (0.029)&   0.140  (0.024)&  0.000 (-0.117)\\
& & $\beta_{1}$& \\\cline{3-3}
1& 0.334 (-0.016)&   0.308 (-0.154)&  0.327 (-0.115)\\
5& -0.014 (-0.014)& 0.040 (-0.040)& -0.029 (0.002)\\
10& 0.030 (-0.001)&  -0.030 (-0.048)&  0.020 (0.021)\\
20& 0.050 (-0.034)&   0.020  (0.003)&  0.050 (-0.030)\\
& & $\beta_{2}$& \\\cline{3-3}
1& 0.088 (-0.072)&  0.043 (-0.051)&  0.051 (-0.022)\\
5& 0.073 (-0.029)&   0.005 (-0.067)&  0.058 (-0.012)\\
10&0.022 (-0.034)&  0.020 (-0.011)&  0.015 (-0.027)\\
20&-0.010 (-0.041)&  0.003  (0.040)& -0.013 (-0.037)\\
& & $\gamma_{1}$& \\\cline{3-3}
1& 0.111 (0.072)&  0.032 (-0.049)&  0.024 (-0.015)\\
5& 0.029 (0.188)& -0.010  (0.065)&  0.011 (0.029)\\
10&0.000 (0.115)& -0.020  (0.034)& -0.008 (-0.068)\\
20&-0.008 (0.191)& -0.008  (0.118)& -0.013 (-0.011)\\
& & $\gamma_{2}$& \\\cline{3-3}
1& 0.244 (-0.044)&  0.114 (-0.212)&  0.151 (-0.123)\\
5& 0.000 (-0.054)&  0.040 (-0.081)& -0.004 (-0.085)\\
10&0.013 (0.178)&  0.060  (0.054)& 0.008 (-0.011)\\
20&-0.013 (0.019)&  0.003 (-0.013)& -0.015 (-0.025)\\
\hline
\end{tabular}
\end{table}

\begin{table}
\caption{Relative bias for regression coefficients (relative bias of standard errors in parenthesis). Logistic model of equation (1) with $\sigma_{j_{1}}^{2}=0.25$ and $\sigma_{j_{2}}^{2}=0.01$. Complete cross-classification with $N_{1}=N_{2}=10$ and varying number of observations per cell $n$.}\label{table:3}
\centering
\begin{tabular}{|c|ccc|}
\hline
$n$ & INLA Ga(0.001,0.001)& INLA Ga(0.5,0.003737)& MLLA\\
\hline
& & $\alpha$& \\\cline{3-3}
1&  0.326 (0.076)& 0.118 (-0.103)&  0.160 (-0.042)\\
5&  0.120 (0.188)&   0.320 (-0.039)&  0.100 (-0.027)\\
10& -0.040 (0.109)&  0.090 (-0.041)& -0.050 (-0.140)\\
20& -0.060 (0.023)& 0.020 (-0.005)& -0.070 (-0.164)\\
& & $\beta_{1}$& \\\cline{3-3}
1&0.383 (-0.120)& 0.393 (-0.072)&  0.300 (-0.059)\\
5&0.010 (-0.015)& 0.000 (-0.066)& -0.010 (0.001)\\
10&0.050 (-0.023)& -0.010 (-0.027)&  0.040 (-0.015)\\
20&0.020 (-0.026)&  0.000 (-0.094)&  0.010 (-0.023)\\
& & $\beta_{2}$& \\\cline{3-3}
1&0.090 (-0.084)&  0.029 (-0.072)&  0.027 (-0.031)\\
5&0.077 (0.010)&  0.013 (-0.040)&  0.062 (0.025)\\
10&0.025 (0.006)& -0.010  (0.004)&  0.015 (0.014)\\
20&-0.005 (-0.058)&  0.005 (-0.032)& -0.008 (-0.054)\\
& & $\gamma_{1}$& \\\cline{3-3}
1&0.166 (0.013)&  0.099 (-0.168)&  0.077 (-0.083)\\
5&0.008 (0.132)&  -0.095 (-0.051)& -0.015 (-0.081)\\
10&0.027 (0.107)&  -0.045 (-0.022)&  0.017 (-0.151)\\
20&-0.013 (0.073)& 0.025  (0.014)& -0.015 (-0.121)\\
& & $\gamma_{2}$& \\\cline{3-3}
1& 0.145 (0.050)&  0.108 (-0.053)&  0.045 (0.010)\\
5&-0.013 (-0.073)& 0.047  (0.023)& -0.023 (-0.107)\\
10&  0.050 (0.061)&  0.070 (-0.060)&  0.040 (-0.186)\\
20&  -0.015 (0.021)& -0.015  (0.011)& -0.020 (-0.001)\\
\hline
\end{tabular}
\end{table}

\subsection{Comparing scenarios with different degrees of cross-classification}
In Tables \ref{table:partial1}, \ref{table:partial2bis} and \ref{table:partial3} we report the relative bias for the fixed effects of model (1), with variance components equal to, respectively, (0.01,0.01), (0.01,0.25) and (0.25,0.01), obtained in scenarios with different degrees of cross-classification and varying number of observations per cell $n$.

\begin{table}
\caption{Relative bias for regression coefficients (relative bias of standard errors in parenthesis). Logistic model of equation (1) with $\sigma_{j_{1}}^{2}=\sigma_{j_{2}}^{2}=0.01$. Structures with different degrees of cross-classification (10 feeders and varying number of receivers), and varying number of observations per cell $n$.}\label{table:partial1}
\centering
\begin{tabular}{|cc|ccc|}
\hline
$n$ & Receivers& INLA Ga(0.001,0.001)& INLA Ga(0.5,0.003737)& MLLA\\
\hline
&& & $\alpha$& \\\cline{4-4}
50& 2& 0.080 (0.173)& 0.050 (0.062)& 0.070 (0.013)\\
20& 5& 0.040 (0.165)& 0.070 (0.544)& 0.030 (0.021)\\
10& 10& 0.010 (0.161)& -0.030 (0.142)& 0.000 (-0.029)\\
&& & $\beta_{1}$& \\\cline{4-4}
50& 2&-0.020 (0.004)& 0.020 (0.000)& -0.020 (-0.002)\\
20& 5& 0.010 (0.041)&  0.020 (0.026)&  0.000 (0.050)\\
10& 10&0.000 (-0.026)& -0.030 (0.010)& -0.010 (-0.018)\\
&& & $\beta_{2}$& \\\cline{4-4}
50& 2&-0.018 (-0.010)&  0.005  (0.058)& -0.025 (-0.002)\\
20& 5& 0.027 (-0.046)&  0.015 (-0.019)&  0.017 (-0.038)\\
10& 10&0.008 (0.073)&  0.027 (-0.008)&  0.000 (0.084)\\
&& & $\gamma_{1}$& \\\cline{4-4}
50& 2& 0.015 (0.166)&  0.008 (0.030)&  0.005 (-0.036)\\
20& 5&-0.010 (0.206)& -0.028 (0.441)& -0.020 (0.023)\\
10& 10& 0.020 (0.138) &  0.005 (0.120)&  0.013 (-0.051)\\
&& & $\gamma_{2}$& \\\cline{4-4}
50& 2&0.015 (-0.008)&  -0.005 (-0.015)& 0.008 (0.000)\\
20& 5& 0.013 (0.017)& 0.032 (-0.050)& 0.008 (0.018)\\
10& 10&0.020 (0.152)& -0.010  (0.124)& 0.010 (-0.045)\\
\hline
\end{tabular}
\end{table}

\begin{table}
\caption{Relative bias for regression coefficients (relative bias of standard errors in parenthesis). Logistic model of equation (1) with $\sigma_{j_{1}}^{2}=0.01$ and $\sigma_{j_{2}}^{2}=0.25$. Structures with different degrees of cross-classification (10 feeders and varying number of receivers), and varying number of observations per cell $n$.}\label{table:partial2bis}
\centering
\begin{tabular}{|cc|ccc|}
\hline
$n$ & Receivers& INLA Ga(0.001,0.001)& INLA Ga(0.5,0.003737)& MLLA\\
\hline
&& & $\alpha$& \\\cline{4-4}
50& 2& 0.170 (0.142)& 0.140 (0.047)&  0.170 (-0.012)\\
20& 5& 0.050 (0.084)& 0.100 (0.140)&  0.040 (-0.043)\\
10& 10& 0.000 (0.176)& 0.170 (0.042) & -0.010 (0.018) \\
&& & $\beta_{1}$& \\\cline{4-4}
50& 2&-0.050 (0.012)& 0.010  (0.006)& -0.050 (0.019)\\
20& 5& -0.050 (-0.009)& 0.291 (0.389)& -0.050 (-0.004)\\
10& 10&0.030 (-0.001)&  -0.030 (-0.048)&  0.020 (0.021)\\
&& &$\beta_{2}$& \\\cline{4-4}
50& 2&-0.013 (-0.021)&  0.003  (0.062)& -0.020 (-0.013)\\
20& 5&0.025 (-0.067)&  0.102 (0.247)&  0.020 (-0.060)\\
10& 10&0.022 (-0.034)&   0.020 (-0.011)&  0.015 (-0.027)\\
&& &$\gamma_{1}$& \\\cline{4-4}
50& 2&-0.008 (0.239)&-0.028 (0.005)& -0.020 (0.049)\\
20& 5&-0.008 (0.152)& 0.071 (0.424)& -0.015 (-0.005)\\
10& 10&0.000 (0.115)& -0.020 (0.034)& -0.008 (-0.068)\\
&& &$\gamma_{2}$& \\\cline{4-4}
50& 2&0.015 (-0.016)& -0.003 (-0.002)& 0.008 (-0.009)\\
20& 5&0.013 (0.055)& 0.409 (0.144)& 0.008 (0.062)\\
10& 10&0.013 (0.178)& 0.060  (0.054)& 0.008 (-0.011)\\
\hline
\end{tabular}
\end{table}

\begin{table}
\caption{Relative bias for regression coefficients (relative bias of standard errors in parenthesis). Logistic model of equation (1) with $\sigma_{j_{1}}^{2}=0.25$ and $\sigma_{j_{2}}^{2}=0.01$. Structures with different degrees of cross-classification (10 feeders and varying number of receivers), and varying number of observations per cell $n$.}\label{table:partial3}
\centering
\begin{tabular}{|cc|ccc|}
\hline
$n$ & Receivers& INLA Ga(0.001,0.001)& INLA Ga(0.5,0.003737)& MLLA\\
\hline
&& & $\alpha$& \\\cline{4-4}
50& 2& 0.080 (0.154)&  0.100 (-0.013)&  0.070 (-0.038)\\
20& 5& -0.050 (-0.006)&  0.070  (0.239)& -0.060 (-0.167)\\
10& 10& -0.040 (0.109)& 0.090 (-0.041)& -0.050 (-0.140)\\
&& & $\beta_{1}$ & \\\cline{4-4}
50& 2&0.000 (-0.006)& 0.010 (-0.010)&  0.000 (0.000)\\
20& 5&-0.020 (0.032)& 0.020 (-0.058)& -0.030 (0.040)\\
10& 10&0.050 (-0.023)& -0.010 (-0.027)&  0.040 (-0.015)\\
&& &$\beta_{2}$& \\\cline{4-4}
50& 2&-0.010 (-0.021)& 0.027 (0.056)& -0.015 (-0.014)\\
20& 5&0.017 (-0.033)&  0.000 (0.026)&  0.010 (-0.026)\\
10& 10&0.025 (0.006)& -0.010 (0.004)&  0.015 (0.014)\\
&& &$\gamma_{1}$& \\\cline{4-4}
50& 2&0.022 (0.107)&  -0.063  (0.025)&  0.015 (-0.090)\\
20& 5&-0.008 (0.067)& -0.045  (0.157)& -0.015 (-0.122)\\
10& 10&0.027 (0.107)&  -0.045 (-0.022)&  0.017 (-0.151)\\
&& &$\gamma_{2}$& \\\cline{4-4}
50& 2&0.020 (-0.028)& 0.010  (0.003)& 0.013 (-0.013)\\
20& 5&0.025 (0.001)&  0.042  (0.008)& 0.020 (0.009)\\
10& 10&0.050 (0.061)& 0.042  (0.008)& 0.040 (-0.186)\\
\hline
\end{tabular}
\end{table}

\subsection{Assessing the asymptotic behaviour}
In Tables \ref{table:a1}, \ref{table:a2} and \ref{table:a3} we present the relative bias for the fixed effects of model (1), with variance components equal to, respectively, (0.01,0.01), (0.01,0.25) and (0.25,0.01), obtained in scenarios with complete cross-classification, increasing number of clusters per classification and $n=10$ observations per cell.

\begin{table}
\caption{Relative bias for regression coefficients (relative bias of standard errors in parenthesis). Logistic model of equation (1) with $\sigma_{j_{1}}^{2}=\sigma_{j_{2}}^{2}=0.01$. Complete cross-classification with varying number of clusters per classification $N_{1}=N_{2}$ and $n=10$ observations per cell.}\label{table:a1}
\centering
\begin{tabular}{|c|ccc|}
\hline
$N_{1}=N_{2}$ & INLA Ga(0.001,0.001)& INLA Ga(0.5,0.003737)& MLLA\\
\hline
& & $\alpha$& \\\cline{3-3}
10& 0.010 (0.161)& -0.030  (0.142)&  0.000 (-0.029)\\
20& -0.060 (0.031)& -0.040  (0.010)& -0.060 (-0.059)\\
50& -0.010 (0.015)& 0.040  (0.003)& -0.010 (-0.010)\\
80& 0.000 (-0.031)&  0.000 (-0.039)&  0.000 (-0.031)\\
& &$\beta_{1}$& \\\cline{3-3}
10&0.000 (-0.026)& -0.030  (0.010)& -0.010 (-0.018)\\
20&0.020 (0.001)& 0.030 (-0.013)&  0.010 (0.003)\\
50&0.000 (-0.015)& 0.000  (0.018)&  0.000 (-0.015)\\
80& 0.000 (-0.023) & 0.000 (-0.030)&  0.000 (-0.023)\\
& & $\beta_{2}$& \\\cline{3-3}
10&0.008 (0.073)& 0.027 (-0.008)&  0.000 (0.084)\\
20&0.000 (-0.032)& 0.010 (-0.039)& -0.003 (-0.031)\\
50&0.005 (0.009)& -0.005 (-0.066)&  0.005 (0.009)\\
80& 0.000 (-0.047)&  0.000  (0.033)&  0.000 (-0.046) \\
& & $\gamma_{1}$& \\\cline{3-3}
10& 0.020 (0.138)& 0.005 (0.120)& 0.013 (-0.051)\\
20&0.022 (0.105)& 0.005  (0.005)& 0.020 (-0.015)\\
50&0.000 (0.048)&  -0.010 (0.019)&  0.000 (0.023)\\
80& 0.008 (-0.023)& 0.005 (0.013)& 0.008 (-0.057)\\
& &$\gamma_{2}$& \\\cline{3-3}
10&0.020 (0.152)& -0.010  (0.124)& 0.010 (-0.045)\\
20&0.008 (-0.041)&  0.008  (0.009)& 0.005 (-0.088)\\
50&0.003 (0.057)& 0.003 (-0.055)& 0.003 (0.058)\\
80&0.000 (-0.078)& 0.000  (0.011)& 0.000 (-0.078)\\
\hline
\end{tabular}
\end{table}

\begin{table}
\caption{Relative bias for regression coefficients (relative bias of standard errors in parenthesis). Logistic model of equation (1) with $\sigma_{j_{1}}^{2}=0.01$ and $\sigma_{j_{2}}^{2}=0.25$. Complete cross-classification with varying number of clusters per classification $N_{1}=N_{2}$ and $n=10$ observations per cell.}\label{table:a2}
\centering
\begin{tabular}{|c|ccc|}
\hline
$N_{1}=N_{2}$ & INLA Ga(0.001,0.001)& INLA Ga(0.5,0.003737)& MLLA\\
\hline
& & $\alpha$& \\\cline{3-3}
10&0.000 (0.176)&   0.170  (0.042)& -0.010 (0.018)\\
20&-0.020 (0.037)& -0.030 (-0.015)& -0.030 (-0.031)\\
50&0.010 (-0.032)&    0.020 (-0.010)&  0.010 (-0.056)\\
80&0.010 (-0.008)&  -0.010 (-0.019)&  0.010 (-0.024)\\
& &$\beta_{1}$& \\\cline{3-3}
10&0.030 (-0.001)&  -0.030 (-0.048)&  0.020 (0.021)\\
20&0.010 (0.045)&  0.010  (0.027)&  0.010 (0.047)\\
50&-0.010 (-0.038)& 0.000  (0.027)& -0.010 (-0.038)\\
80&0.000 (0.078)& 0.010  (0.071)&  0.000 (0.078)\\
& &$\beta_{2}$& \\\cline{3-3}
10&0.022 (-0.034)&  0.020 (-0.011)&  0.015 (-0.027)\\
20&-0.003 (0.023)& 0.003 (-0.027)& -0.005 (0.024)\\
50&0.008 (0.027)&  0.000 (-0.025)&  0.008 (0.027)\\
80& 0.000 (-0.031)& 0.003  (0.071)& 0.000 (-0.031)\\
& &$\gamma_{1}$& \\\cline{3-3}
10&0.000 (0.115)&  -0.020  (0.034)& -0.008 (-0.068)\\
20&0.015 (0.085)&  0.015  (0.047)&  0.013 (-0.020)\\
50&0.003 (0.052)& -0.008  (0.010)&  0.003 (0.027)\\
80&0.008 (0.010)&  0.005 (-0.004)&  0.008 (-0.025)\\
& & $\gamma_{2}$& \\\cline{3-3}
10&0.013 (0.178)& 0.060  (0.054)&0.008 (-0.011)\\
20&0.017 (-0.005)& -0.003  (0.074)& 0.015 (-0.036)\\
50&0.003 (0.009)& 0.003 (-0.058)& 0.003 (0.009)\\
80&0.000 (-0.034)& 0.000  (0.058)& 0.000 (-0.034)\\
\hline
\end{tabular}
\end{table}

\begin{table}
\caption{Relative bias for regression coefficients (relative bias of standard errors in parenthesis). Logistic model of equation (1) with $\sigma_{j_{1}}^{2}=0.25$ and $\sigma_{j_{2}}^{2}=0.01$. Complete cross-classification with varying number of clusters per classification $N_{1}=N_{2}$ and $n=10$ observations per cell.}\label{table:a3}
\centering
\begin{tabular}{|c|ccc|}
\hline
$N_{1}=N_{2}$ & INLA Ga(0.001,0.001)& INLA Ga(0.5,0.003737)& MLLA\\
\hline
& & $\alpha$& \\\cline{3-3}
10&-0.040 (0.109)& 0.090 (-0.041)& -0.050 (-0.140)\\
20&-0.100 (0.015)& -0.080 (-0.026)& -0.110 (-0.070)\\
50& -0.040 (0.033)& 0.110 (-0.008)& -0.040 (-0.001)\\
80& -0.020 (-0.014)& -0.030 (-0.042)& -0.020 (-0.040)\\
& &$\beta_{1}$& \\\cline{3-3}
10&0.050 (-0.023)&  -0.010 (-0.027)&  0.040 (-0.015)\\
20&0.010 (0.010)&  0.020 (-0.022)&  0.010 (0.012)\\
50&-0.010 (-0.064)& 0.000 (-0.005)& -0.010 (-0.064)\\
80&0.010 (0.076)&  0.000  (0.015)& 0.010 (0.076)\\
& &$\beta_{2}$& \\\cline{3-3}
10&0.025 (0.006)&  -0.010  (0.004)&  0.015 (0.014)\\
20&-0.003 (0.050)& 0.010 (-0.017)& -0.005 (0.053)\\
50&0.003 (-0.009)&  0.000 (-0.013)&  0.003 (-0.009)\\
80&0.000 (0.051)& 0.000 (-0.019)& 0.000 (0.052)\\
& &$\gamma_{1}$& \\\cline{3-3}
10&0.027 (0.107)&  -0.045 (-0.022)& 0.017 (-0.151)\\
20&0.022 (0.037)& 0.017  (0.006)& 0.020 (-0.052)\\
50&0.017 (0.040)& -0.050  (0.011)& 0.017 (0.006)\\
80&0.032 (-0.006)& 0.020 (-0.013)& 0.032 (-0.023)\\
& &$\gamma_{2}$& \\\cline{3-3}
10&0.050 (0.061)&  0.070 (-0.060)& 0.040 (-0.186)\\
20&0.003 (0.015)& -0.015  (0.024)& 0.000 (0.002)\\
50&0.003 (-0.040)& 0.005 (-0.044)& 0.003 (-0.040)\\
80&0.000 (-0.006)&  0.003  (0.060)& 0.000 (-0.006)\\
\hline
\end{tabular}
\end{table}

\section{Results when both variance components are $1.00$}
In this section we compare the performance of the three considered estimators when the magnitude of both variance components is $1.00$. Although random effects of such size are rarely found in applications, it is noteworthy to check whether the behaviour of INLA is coherent with the findings of the paper.

\begin{table}
\caption{Relative bias for regression coefficients (relative bias of standard errors in parenthesis). Logistic model of equation (1) with $\sigma_{j_{1}}^{2}=\sigma_{j_{2}}^{2}=1.00$. Complete cross-classification with $N_{1}=N_{2}=10$ and varying number of observations per cell $n$.}\label{table:6}
\centering
\begin{tabular}{|c|ccc|}
\hline
$n$ & INLA Ga(0.001,0.001)& INLA Ga(0.5,0.003737)& MLLA\\
\hline
& & $\alpha$& \\\cline{3-3}
1&0.118 (-0.083)&   0.176  (-0.265)   & -0.220 (-0.152)\\
5&-0.255 (-0.019)&  0.805  (-0.084)      & -0.097 (-0.139)\\
10&-0.066 (0.006)&  -0.776 (-0.016)      & -0.100 (-0.101)\\
20&-0.253 (-0.034)&   0.188 (-0.040)       & 0.080 (-0.148)\\
& & $\beta_{1}$& \\\cline{3-3}
1&0.037 (-0.123)&   -0.167  (-0.043)  & 0.070 (-0.108)\\
5&0.025 (0.015)&    0.014  (-0.065)      & 0.046 (0.004)\\
10&0.065 (0.016)& 0.039  (-0.008)     & 0.070 (0.002)\\
20&0.006 (0.015)&    -0.003  (-0.009)    &0.030 (0.036)\\
& & $\beta_{2}$& \\\cline{3-3}
1&0.003 (-0.090)&   -0.124 (0.003)    &  0.115 (-0.098)\\
5&0.089 (-0.037)&   0.016  (0.041)     &  0.049 (-0.048)\\
10&-0.002 (-0.023)&    -0.007  (-0.014)    & -0.008 (-0.004)\\
20&0.016 (0.001)&    -0.007 (-0.022)   &0.003 (0.017)\\
& & $\gamma_{1}$& \\\cline{3-3}
1&0.053 (-0.050)&    -0.200   (-0.282)   &  0.010 (-0.171)\\
5& -0.056 (-0.053)&   -0.271  (-0.097)    &  0.007 (-0.153)\\
10&-0.033 (-0.062)&   0.130  (-0.058)   & -0.008 (-0.186)\\
20&0.062 (-0.054)&   0.013 (-0.019)        & -0.050 (-0.181)\\
& & $\gamma_{2}$& \\\cline{3-3}
1&0.179 (-0.170)&   0.005  (-0.339) &  0.277 (-0.217)\\
5&0.059 (-0.085)&    -0.109  (-0.064) &  0.028 (-0.080)\\
10&0.004 (-0.042)&  0.103   (-0.056)    & -0.045 (-0.179)\\
20&0.011 (0.025)&    -0.022  (0.014)  &  0.015 (0.006)\\
\hline
\end{tabular}
\end{table}

\begin{table}
\caption{Relative bias for regression coefficients (relative bias of standard errors in parenthesis). Logistic model of equation (1) with $\sigma_{j_{1}}^{2}=\sigma_{j_{2}}^{2}=1.00$. Structures with different degrees of cross-classification (10 feeders and varying number of receivers), and varying number of observations per cell $n$.}\label{table:partial33}
\centering
\begin{tabular}{|cc|ccc|}
\hline
$n$ & Receivers& INLA Ga(0.001,0.001)& INLA Ga(0.5,0.003737)& MLLA\\
\hline
&& & $\alpha$& \\\cline{4-4}
50& 2& 0.389 (0.006)&   0.386 (-0.023)  &  0.380 (-0.046)\\
20& 5& -0.253 (-0.084)&  0.337 (0.174)   &  0.020 (-0.107)\\
10& 10& -0.066  (0.006)&  -0.776  (-0.016)   & -0.100 (-0.101)\\
&& &$\beta_{1}$& \\\cline{4-4}
50& 2&-0.023 (0.020)&   -0.029 (0.001)    & -0.030 (0.036)\\
20& 5&-0.019 (-0.051)&    0.002 (-0.030)    & -0.050 (-0.035)\\
10& 10& 0.065  (0.016)& 0.039  (-0.008)   &  0.070 (0.002)\\
&& &$\beta_{2}$& \\\cline{4-4}
50& 2& 0.005 (-0.002)&   0.025 (0.008) & -0.013 (-0.011)\\
20& 5&0.000 (-0.053)&  0.008  (-0.011)      &  0.008 (-0.055)\\
10& 10&-0.002 (-0.023)&  -0.007  (-0.014)    & -0.008 (-0.004)\\
&& &$\gamma_{1}$& \\\cline{4-4}
50& 2&-0.020 (-0.002)&  -0.206  (-0.013)     & -0.033 (-0.078)\\
20& 5&0.040  (-0.011)&  -0.115  (0.080)    & -0.043 (-0.094)\\
10& 10&-0.033 (-0.062)& 0.130  (-0.058)   & -0.008 (-0.186)\\
&& &$\gamma_{2}$& \\\cline{4-4}
50& 2&0.037 (-0.008)&    0.004 (0.006)    &  0.027 (-0.012)\\
20& 5&0.009 (0.005)&  0.025 (0.030)    & -0.003 (-0.021)\\
10& 10&0.004 (-0.042)& 0.103  (-0.056)  & -0.045 (-0.179)\\
\hline
\end{tabular}
\end{table}

\begin{table}
\caption{Relative bias for regression coefficients (relative bias of standard errors in parenthesis). Logistic model of equation (1) with $\sigma_{j_{1}}^{2}=\sigma_{j_{2}}^{2}=1.00$. Complete cross-classification with varying number of clusters per classification $N_{1}=N_{2}$ and $n=10$ observations per cell.}\label{table:a10}
\centering
\begin{tabular}{|c|ccc|}
\hline
$N_{1}=N_{2}$ & INLA Ga(0.001,0.001)& INLA Ga(0.5,0.003737)& MLLA\\
\hline
& & $\alpha$& \\\cline{3-3}
10&-0.066  (0.006)   &   -0.776  (-0.016)     & -0.100 (-0.101)\\
20&-0.150 (0.032)&   0.019 (0.014)  & -0.120 (-0.025)\\
50&-0.070 (-0.024)&  0.080 (0.025)     & -0.070 (-0.048)\\
80&-0.020 (-0.005)& -0.070 (-0.058)      & -0.020 (-0.021)\\
& &$\beta_{1}$& \\\cline{3-3}
10&0.065  (0.016)   &    0.039  (-0.008)         & 0.070 (0.002)\\
20&0.008 (0.011)&  -0.014  (0.003)    & 0.010 (0.029)\\
50&0.000 (0.006)&   0.000 (-0.038)    & 0.000 (0.006)\\
80&0.000 (0.036)&  0.010  (0.065)     & 0.000 (0.036)\\
& &$\beta_{2}$& \\\cline{3-3}
10&-0.002 (-0.023)   &   -0.007  (-0.014)     & -0.008 (-0.004)\\
20&0.003 (0.024)&  -0.006  (0.001)    &  0.000 (0.025)\\
50&0.003 (0.011)&  -0.003 (-0.046)   &  0.003 (0.011)\\
80&-0.003 (-0.048)& 0.000 (0.018)    & -0.003 (-0.048)\\
& &$\gamma_{1}$& \\\cline{3-3}
10&-0.033 (-0.062)   &    0.130  (-0.058)       & -0.008 (-0.186)\\
20& 0.023 (0.022)&   0.049 (-0.011)       &  0.027 (-0.043)\\
50&-0.020 (-0.024)&  -0.020  (-0.009)      & -0.020 (-0.051)\\
80&0.060 (-0.003)&  0.043  (-0.001)      &  0.060 (-0.021)\\
& &$\gamma_{2}$& \\\cline{3-3}
10& 0.004 (-0.042)   &    0.103  (-0.056)   & -0.045 (-0.179)\\
20&-0.004 (-0.018)&  0.010  (-0.003)      & -0.008 (-0.019)\\
50&-0.003 (0.011)&  0.000  (0.031)      & -0.003 (0.011)\\
80&-0.003 (-0.025)& 0.000 (0.000)        & -0.003 (-0.025)\\
\hline
\end{tabular}
\end{table}

\begin{table}
\caption{Percentage of extreme estimates out of the 500 replicates. Complete cross-classification with $N_{1}=N_{2}=10$, varying number of observations per cell $n$ and $\sigma_{j_{1}}^{2}=\sigma_{j_{2}}^{2}=1.00$} \label{table:anomalie4}
\centering
\begin{tabular}{|c|ccc|}
\hline
 & INLA Ga(0.001,0.001)& INLA Ga(0.5,0.003737)& MLLA\\
\hline
  &\% $\hat{\sigma}_{j_{1}}^{2}> 2$ \& \% $\hat{\sigma}_{j_{2}}^{2}> 2$ &\% $\hat{\sigma}_{j_{1}}^{2}> 2$ \& \% $\hat{\sigma}_{j_{2}}^{2}> 2$ &\% $\hat{\sigma}_{j_{1}}^{2}=0$ \& \% $\hat{\sigma}_{j_{2}}^{2}=0$ \\
\hline
1& 37.2  $\;\;\;\;\;\;$    35.8 &    25.8$\;\;\;\;\;\;$      26.2 &     3.6   $\;\;\;\;\;\;$    3.0\\
5& 15.6  $\;\;\;\;\;\;$    15.6 &     12.6  $\;\;\;\;\;\;$     8.2  &    0.0   $\;\;\;\;\;\;$    0.0\\
10& 19.8  $\;\;\;\;\;\;$    14.8  &    11.4  $\;\;\;\;\;\;$     5.6   &      0.0    $\;\;\;\;\;\;$      0.0\\
20& 12.8  $\;\;\;\;\;\;$    13.0 &     7.0  $\;\;\;\;\;\;$     7.6   &      0.0     $\;\;\;\;\;\;$     0.0\\\hline
\end{tabular}
\end{table}

\begin{table}
\caption{Percentage of extreme estimates out of the 500 replicates. Structures with different degree of cross-classification, varying number of observations per cell $n$ and $\sigma_{j_{1}}^{2}=\sigma_{j_{2}}^{2}=1.00$} \label{table:anomalie5}
\centering
\begin{tabular}{|c|ccc|}
\hline
 & INLA Ga(0.001,0.001)& INLA Ga(0.5,0.003737)& MLLA\\
\hline
  &\% $\hat{\sigma}_{j_{1}}^{2}> 2$ \& \% $\hat{\sigma}_{j_{2}}^{2}> 2$ &\% $\hat{\sigma}_{j_{1}}^{2}> 2$ \& \% $\hat{\sigma}_{j_{2}}^{2}> 2$ &\% $\hat{\sigma}_{j_{1}}^{2}=0$ \& \% $\hat{\sigma}_{j_{2}}^{2}=0$ \\
\hline
2& 20.0  $\;\;\;\;\;\;$    16.6&     10.6  $\;\;\;\;\;\;$     8.4  &     0.0     $\;\;\;\;\;\;$     0.0\\
5& 17.8  $\;\;\;\;\;\;$    17.8 &     8.6   $\;\;\;\;\;\;$    9.4  &       0.0    $\;\;\;\;\;\;$      0.0\\
10& 19.8   $\;\;\;\;\;\;$   14.8  &    11.4  $\;\;\;\;\;\;$     5.6  &       0.0    $\;\;\;\;\;\;$      0.0\\\hline
\end{tabular}
\end{table}

\begin{table}
\caption{Percentage of extreme estimates out of the 500 replicates. Complete cross-classification with varying number of clusters per classification $N_{1}=N_{2}$, $n=10$ and $\sigma_{j_{1}}^{2}=\sigma_{j_{2}}^{2}=1.00$} \label{table:anomalie6}
\centering
\begin{tabular}{|c|ccc|}
\hline
 & INLA Ga(0.001,0.001)& INLA Ga(0.5,0.003737)& MLLA\\
\hline
  &\% $\hat{\sigma}_{j_{1}}^{2}> 2$ \& \% $\hat{\sigma}_{j_{2}}^{2}> 2$ &\% $\hat{\sigma}_{j_{1}}^{2}> 2$ \& \% $\hat{\sigma}_{j_{2}}^{2}> 2$ &\% $\hat{\sigma}_{j_{1}}^{2}=0$ \& \% $\hat{\sigma}_{j_{2}}^{2}=0$ \\
\hline
10& 19.8   $\;\;\;\;\;\;$   14.8 &    11.4   $\;\;\;\;\;\;$    5.6   &   0.0     $\;\;\;\;\;\;$     0.0\\
20& 2.2  $\;\;\;\;\;\;$    1.6 &    1.0   $\;\;\;\;\;\;$    0.8 &  0.0   $\;\;\;\;\;\;$     0.0\\
50& 0.0  $\;\;\;\;\;\;$    0.0  &    0.0  $\;\;\;\;\;\;$     0.0  &       0.0     $\;\;\;\;\;\;$     0.0\\
80& 0.0   $\;\;\;\;\;\;$   0.0  &    0.0 $\;\;\;\;\;\;$      0.0  &       0.0    $\;\;\;\;\;\;$     0.0\\
\hline
\end{tabular}
\end{table}

\begin{figure}                               
\centerline{\includegraphics[scale=0.3]{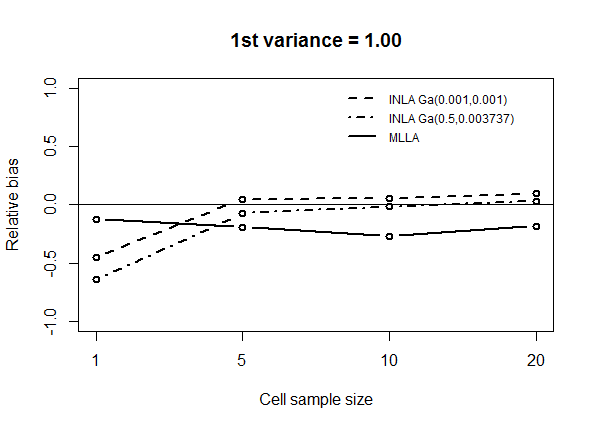} \includegraphics[scale=0.3]{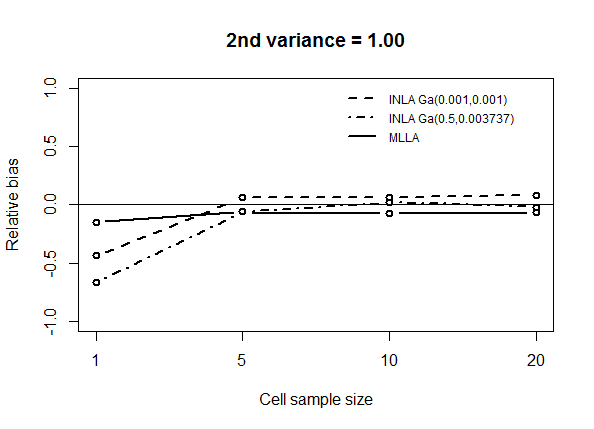} }
\centerline{\includegraphics[scale=0.3]{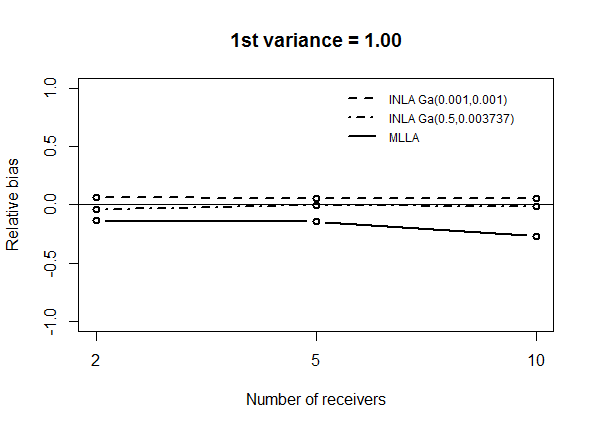}
\includegraphics[scale=0.3]{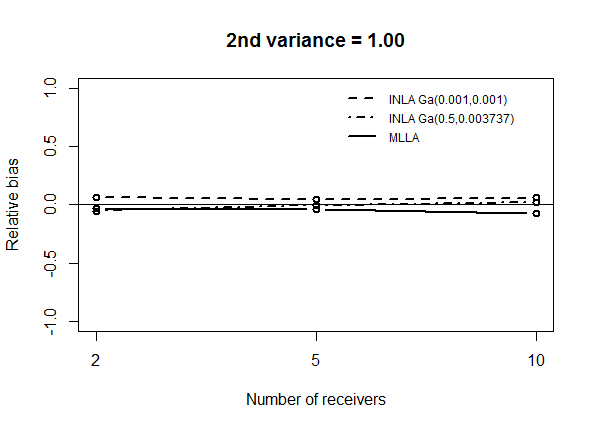}}
\centerline{
\includegraphics[scale=0.3]{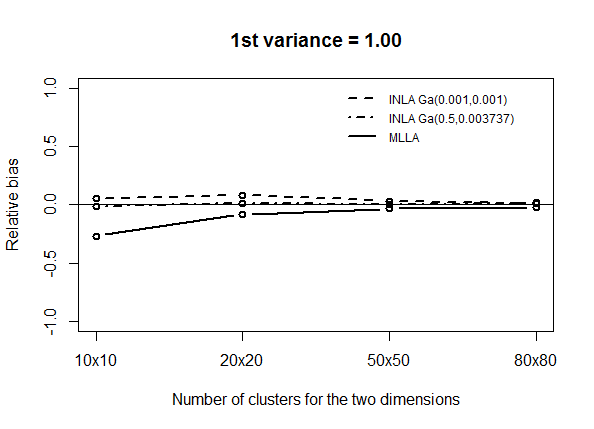}
\includegraphics[scale=0.3]{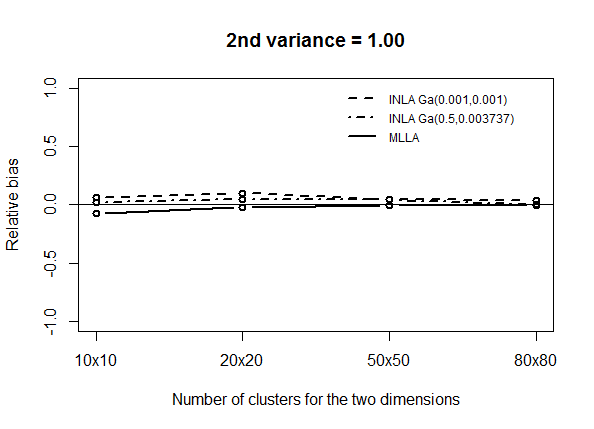}}
\caption{Relative bias for the variance components of the logistic model of equation (1). First row: complete cross-classification with varying numbers of clusters per dimension $N_{1} \times N_{2}$ and $n=10$ observations per cell. Second row: structures with different degrees of cross-classification (10 feeders and varying number of receivers). The cell sample size $n$ is set on the basis of the number of receivers to ensure a total sample size of 1000. Third row: complete cross-classification with varying numbers of clusters per dimension $N_{1} \times N_{2}$ and $n=10$ observations per cell. Each pair of graphs corresponds to a combination of random effects variances ($\sigma_{j_{1}}^{2}, \sigma_{j_{2}}^{2}$): (1.00, 1.00).}
\label{figure:var1}
\end{figure}

While the patterns of the bias are similar (apart for the intercept $\alpha$) to those observed in scenerios with smaller variances (see Tables \ref{table:6}, \ref{table:partial33}, \ref{table:a10} and Figure \ref{figure:var1}), the seriousness of extreme estimates is reversed with respect to the results presented in the paper (see Tables \ref{table:anomalie4}, \ref{table:anomalie5} and \ref{table:anomalie6}): in scenarios with both variances equal 1.00 this issue affects more INLA than MLLA. Indeed, while the frequentist method gives zero estimates only when the cross-classification matrix is $10\times 10$ and $n=1$, the Bayesian algorithm needs at least 100 clusters to yield no aberrant estimates. Moreover, it is worth to note the difference between the two priors, which may be explained by the fact that Gamma(0.5,0.003737) is weakly informative, whereas Gamma(0.001,0.001) is absolutely flat.
However, it is quite obvious that in these scenarios, INLA is more affected by this issue than MLLA, since the magnitude of variances is close to the specified threshold for INLA ($\hat{\sigma}_{j_{.}}^{2}> 2$) and considerably far from the inferior border of the parameter space. Furthermore, variances equal to 1.00 are rarely found in applications, reducing the severity of this issue. 

\clearpage

\newpage

\section{An anomalous behaviour of INLA with the prior Ga(0.001,0.001)}
\begin{figure}[h]
 \centering{
 \includegraphics[scale=0.5]{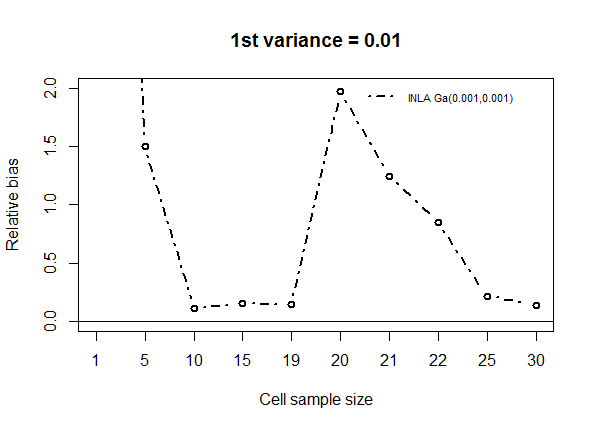}}
 \caption{Relative bias for $\sigma_{j_{1}}^{2}$ of the logistic model of equation (1), estimated with INLA Ga(0.001,0.001). Complete cross-classification with $N_{1}=N_{2}=10$, $\sigma_{j_{1}}^{2}=0.01, \sigma_{j_{2}}^{2}=0.25$ and varying number of observations per cell $n$.}
\label{figure:casoanomalo}
 \end{figure}

In the configuration with variance components $\sigma_{j_{1}}^{2}=0.01$ and $\sigma_{j_{2}}^{2}=0.25$, INLA with the prior Gamma(0.001,0.001) shows an anomalous behaviour because the bias abruptly increases when moving from cell size $n=10$ to $n=20$. This pattern is confirmed by repeating the simulation exercise using a finer grid of values for the cell size (see Figure \ref{figure:casoanomalo}). The large bias is determined by several large estimates, though not aberrant (i.e. not larger than 2), thus the situation is difficult to detect. In real applications, a cautionary approach to limit the bias is to summarize the posterior distribution with the median, which is more robust than the mean (see for example Karim \& Zeger (1992)).

\end{document}